\documentclass[pdflatex,sn-nature]{sn-jnl}% Style for submissions to Nature Portfolio journals

\usepackage{graphicx}
\usepackage{multirow}
\usepackage{amsmath,amssymb,amsfonts}
\usepackage{amsthm}
\usepackage{mathrsfs}
\usepackage[title]{appendix}
\usepackage{xcolor}
\usepackage{textcomp}
\usepackage{manyfoot}
\usepackage{booktabs}
\usepackage{algorithm}
\usepackage{algorithmicx}
\usepackage{algpseudocode}
\usepackage{listings}
\usepackage{subcaption}

\usepackage{braket}
\usepackage{array}
\usepackage{placeins}
\usepackage{anyfontsize}
\usepackage{t1enc}

\usepackage{url}
\usepackage{float}
\usepackage{longtable}
\usepackage{seqsplit}
\usepackage{bookmark}
\usepackage{enumitem}
\newcolumntype{C}[1]{>{\centering\arraybackslash}m{#1}}
\newcolumntype{L}[1]{>{\raggedright\arraybackslash}p{#1}}
\newcommand{\ignore}[1]{}

\begin{document}
\title{Adaptive Reinforcement Learning for Robust Open Quantum System Control: A Multi-Task Framework with Temporal Optimization}

\author*[1,3]{\fnm{Haftu W.} \sur{Fentaw}}\email{haftu.fentaw@ucdconnect.ie}

\author[2,3]{\fnm{Steve} \sur{Campbell}}
\equalcont{These authors contributed equally to this work.}

\author[1,3]{\fnm{Simon} \sur{Caton}}
\equalcont{These authors contributed equally to this work.}

\affil[1]{\orgdiv{School of Computer Science}, \orgname{University College Dublin}, \orgaddress{\street{Belfield}, \city{Dublin 4}, \state{Dublin}, \country{Ireland}}}

\affil[2]{\orgdiv{School of Physics}, \orgname{University College Dublin}, \orgaddress{\street{Belfield}, \city{Dublin 4}, \state{Dublin}, \country{Ireland}}}

\affil[3]{\orgdiv{Centre for Quantum Engineering, Science, and Technology, University College Dublin}, \orgaddress{\street{Belfield}, \city{Dublin 4}, \state{Dublin}, \country{Ireland}}}

\abstract{We present a Multi-task Soft Actor-Critic (SAC) Reinforcement Learning framework designed for open-system quantum control across diverse Hamiltonians, which learns optimal pulse sequences while simultaneously discovering problem-specific evolution time $T$ and number of control pulse segments $N$. Experimental results across 51 Hamiltonian variations demonstrate that the multi-task SAC model is able to generate control pulses that can drive a system, under environment noise, from its initial state to its target state with high fidelities, establishing essential foundations for universal quantum control applicable to realistic noisy quantum devices. Through progressive expansion of the training Hamiltonian set, we investigate if a single multi-task model trained using a given number of sample Hamiltonians can successfully accomplish state-transfer tasks for Hamiltonians drawn from the same Hamiltonian space but not encountered during training. In addition, our Robustness Infidelity Measure (RIM) analysis reveals that SAC trained policies exhibit superior robustness to pulse amplitude perturbations and decoherence rate variations compared to GRAPE-optimized controls.}


\maketitle

\section{Introduction}\label{sec:introduction}

Quantum control theory has evolved from foundational theoretical constructs into practical methodologies that are central to the operation of near-term intermediate-scale quantum (NISQ) devices~\cite{preskill2018quantum}. While early developments primarily focused on idealized closed-system dynamics, realistic quantum hardware inevitably operates as an open-system, continuously interacting with its surrounding environment. This shift from closed to open quantum system introduces fundamental challenges that classical and gradient-based control techniques struggle to address effectively~\cite{weidner2024robustquantumcontrolclosed}.

Gradient Ascent Pulse Engineering (GRAPE)~\cite{khaneja2005optimal} and related gradient-based methods, while achieving high fidelities for quantum control tasks under ideal conditions, suffer from sensitivity to modeling uncertainties, parameter variations, and environmental noise that characterize real quantum hardware~\cite{zhang2025meta}. As a result, control policies optimized under closed-system assumptions can fail to generalize reliably to realistic experimental conditions.

Recent advances in reinforcement learning (RL) for quantum control~\cite{bukov2018reinforcement, niu2019universal, zhang2019reinforcement, bukov2026reinforcement} have demonstrated promising capabilities in discovering effective control pulses and improving robustness to uncertainties. Nevertheless, most existing RL-based approaches focus on individual quantum systems and treat temporal parameters—such as the total evolution time $T$ and the number of discretization steps (pulse segments) $N$—as fixed hyperparameters~\cite{sarma2025designing, bukov2018reinforcement, RLPCA2025}. Moreover, these approaches—including the present work—typically adopt a piecewise-constant pulse parametrization, where the agent specifies independent control amplitudes at each time step. While this is the most common choice in both gradient-based optimal control methods such as GRAPE and RL-based frameworks, recent work has shown that the choice of basis functions for pulse parametrization (e.g., Fourier, sinc, or sigmoid bases) can impact optimization efficiency~\cite{PhysRevA110062608}. The restriction to fixed temporal parameters may limit the optimization landscape, for example by having too many or too few control segments, and prevent the discovery of problem-specific temporal structures that could enhance both control fidelity and robustness.

In practice, quantum devices are subject to decoherence through amplitude damping ($T_1$, energy relaxation) and phase damping ($T_2$, dephasing) processes~\cite{Nielsen_Chuang_2010, manzano_2020}, rendering idealized closed-system models inadequate for realistic control design. The performance gap that arises when applying closed-system trained controllers to open quantum systems is therefore worth exploring. While closed-system controllers may achieve high fidelity, our numerical experiments reveal a substantial degradation in performance when these policies are used in open-system settings. Moreover, this degradation is non-uniform across different Hamiltonians, with certain systems exhibiting heightened sensitivity to environmental noise. We further demonstrate that retraining control policies with explicit open-system dynamics significantly improves both fidelity and robustness. Detailed experimental results supporting these observations are presented in Section~\ref{sec:results_and_discusion}.

Meta-learning~\cite{finn2017model} approaches have been proposed to develop quantum control models that can be trained either across diverse systems for rapid adaptation to specific tasks~\cite{zhang2025meta}, or on a single system with varying parameters, such as coupling coefficients or decoherence levels~\cite{leclerc2026does}. These methods typically involve a two-stage process: a meta-training phase using support and query sets to learn a generalized prior, followed by an adaptation phase to fine-tune the model for a target task. In contrast, we propose a single unified multi-task reinforcement learning framework based on Soft Actor-Critic (SAC)~\cite{haarnoja2018soft, Haarnoja2018SoftAA} that directly synthesizes optimal control policies for a diverse set of Hamiltonians under open-system dynamics. Although task-specific controllers can achieve optimal performance for individual Hamiltonians, training separate models for each Hamiltonian scales poorly as system variability increases. Realistic quantum platforms exhibit variations in Hamiltonian parameters across devices and operating conditions, making per-system training impractical. Our unified multi-task approach provides a scalable alternative by enabling a single control policy to operate across diverse Hamiltonians. For open quantum systems, targeting a benchmark fidelity of $0.95$—a threshold chosen to account for the inherent limitations of certain open-system dynamics, where decoherence imposes hard physical ceilings on achievable fidelity—allows us to demonstrate broad multi-task SAC effectiveness without being limited by system specific constraints. Consequently, this generalized policy serves as a powerful initialization tool. In complex settings where cold-start searches are prohibitively expensive, our model provides a refined baseline that can be further optimized toward the high-fidelity limits required for physical implementations.

Unlike existing methods with rigid temporal constraints, our framework enables adaptive optimization of both the total evolution time $T$ and the number of control segments $N$. In addition to achieving high-fidelity state preparation in the presence of environmental noise, the learned control policies exhibit enhanced robustness to uncertainties such as pulse amplitude fluctuations and variations in decoherence rates. This robustness is quantitatively evaluated using a Robustness Infidelity Measure (RIM)~\cite{PhysRevA.107.032606}, which demonstrates the agent’s ability to sustain performance under stochastic environmental perturbations.

\vspace{1em}
\noindent This paper makes the following contributions:
\begin{itemize}
    \item A unified multi-task SAC based framework capable of learning state transfer across diverse Hamiltonian variations (single-qubit, two-qubit, and three-qubit systems)
     \item A comprehensive robustness analysis comparing multi-task SAC and GRAPE-based control techniques under pulse amplitude perturbations, decoherence rate variations, and combined noise scenarios.
     \item An adaptive temporal optimization strategy demonstrating that learning problem-specific evolution times $T$ and discretization steps $N$ yields improved fidelity and enhanced computational efficiency relative to fixed temporal parameter approaches.
     \item An investigation of the training set diversity required for effective generalization to unseen Hamiltonians, addressing the model’s ability to generalize across previously unseen systems and state transfer tasks.
    \item A systematic comparison of control policies trained on closed versus open quantum system dynamics, quantifying the resulting performance gap and highlighting the necessity of open-system training for practical quantum control.
\end{itemize}

The remainder of this paper is organized as follows. Section~\ref{sec:background_and_related_work} reviews the theoretical background and related work. Section~\ref{sec:methodology_and_experimental_setup} presents the methodology and experimental setup, while Section~\ref{sec:results_and_discusion} discusses the experimental results. Finally, Section~\ref{sec:conclusion} concludes the paper and outlines directions for future research. Detailed specifications of the Hamiltonians are provided in Appendix~\ref{appendix:A}.

%==============================================================================
\section{Background and Related Work} \label{sec:background_and_related_work}
%==============================================================================

\subsection{Quantum System Dynamics}
For an isolated quantum system, the dynamics are governed by the time-dependent Schrödinger equation:
\begin{equation}
    i\hbar\frac{\partial}{\partial t}|\psi(t)\rangle = H(t)|\psi(t)\rangle
    \label{eq:schrodinger}
\end{equation}
where the total Hamiltonian $H(t) = H_0 + \sum_j u_j(t)H_j$ comprises a drift term $H_0$ (intrinsic system dynamics and static terms) and control operators $H_j$ modulated by time-dependent control fields $u_j(t)$. The control fields $u_j(t)$ typically correspond to physical quantities such as microwave pulse amplitudes, laser intensities, or magnetic field strengths that can be shaped to steer the quantum state toward a desired target~\cite{Ian_Rabitz_2003_quantum_control}.

Real quantum systems interact with their environment, leading to decoherence and dissipation. To model the open-system dynamics under such interactions, we employ the Lindblad master equation~\cite{lindblad1976generators,manzano_2020}, which is derived under the Born–Markov and secular approximations~\cite{BreuerPetruccione2002}. Specifically, the Born approximation assumes weak system–bath coupling, such that the environment remains negligibly disturbed and system–bath correlations are minimal. The Markov approximation assumes a memoryless bath, meaning the bath correlation time is much shorter than the system’s characteristic evolution time; as a result, the system’s future evolution depends only on its present state, not on its history. The Secular approximation neglects any fast rotating terms that average to zero over the evolution time. Under these approximations the Lindblad master equation is formulated as: 
\begin{equation}
    \frac{d\rho}{dt} = -i[H(t), \rho] + \sum_k \gamma_k\left(L_k \rho L_k^\dagger - \frac{1}{2}\{L_k^\dagger L_k, \rho\}\right)
    \label{eq:lindblad}
\end{equation}
where $\rho$ is the density matrix, $L_k$ are Lindblad operators describing dissipative channels, and $\gamma_k$ are associated decay rates.

As mentioned above, for quantum systems, the main decoherence mechanisms are amplitude damping and phase damping. 
In this work, we focus on amplitude damping (energy relaxation), where the population decays from the excited state to the ground state described by a time independent Lindblad operator $L_1 = \sqrt{\gamma_1}\sigma_-$~\cite{lindblad1976generators, BreuerPetruccione2002}, where $\gamma_1 = 1/T_1$, is the is the relaxation rate, $T_1$ is relaxation time and $\sigma_-$ is a lowering operator.

In addition to these intrinsic noise channels, quantum systems are subject to systematic errors, including fluctuations in Hamiltonian parameters and variations in control pulse amplitudes~\cite{Ball_2021}. Accordingly, we perform a robustness analysis that considers perturbations in pulse amplitudes, variations in decoherence rates (restricted to amplitude damping), as well as the combined effect of both noise sources.

In order to find optimal control pulses, we must solve either the Schrödinger equation~\eqref{eq:schrodinger} for closed-systems or the Lindblad master equation~\eqref{eq:lindblad} for open-systems. A common numerical approach is the discrete-time approximation, which involves partitioning the total evolution time $T$ into $N$ equal segments. Consequently, both $T$ and $N$ are important parameters that dictate the accuracy and resolution of the simulated time evolution. We utilize the \textit{sesolve} and \textit{mesolve} modules from the QuTiP library~\cite{qutip5} to simulate the time evolution for the closed-system and open-system dynamics. 

\subsection{Hamiltonian Space for Quantum Control} \label{hamiltonian_details}

In this work, we consider single-qubit, two-qubit, and three-qubit quantum systems. We begin by defining the general controlled Hamiltonian for a single-qubit system, which takes the form:
\begin{equation}
    H(t) = \omega_x \sigma_x + \omega_y \sigma_y + \omega_z \sigma_z + u_x(t)\sigma_x + u_y(t)\sigma_y + u_z(t)\sigma_z
    \label{eq:single_qubit_ham}
\end{equation}
where $\omega_{x,y,z}$ represent static field strengths, $u_{x,y,z}(t)$ are time-dependent control amplitudes, and $\sigma_{x,y,z}$ are Pauli matrices--the fundamental 2×2 matrices that generate single-qubit rotations about the $x$, $y$, and $z$ axes of the Bloch sphere.

We note that from a purely physical perspective, permuting Pauli operators (e.g., substituting $\sigma_x$ with $\sigma_y$ in a Hamiltonian) amounts to a rotation of the reference frame and does not alter the underlying physics. However, these formulations are mathematically distinct from the perspective of the SAC agent. Because the agent's observation space is anchored to a fixed computational basis, $\sigma_x$, $\sigma_y$, $\sigma_z$ each produce different numerical density matrix trajectories. This helps the agent generalize across Hamiltonians that might be physically equivalent but mathematically distinct in the chosen basis.

We first focus our experimentation on single-qubit systems using a standard and well studied driven-qubit Hamiltonian of the form $ H = \sigma_x + u(t) \sigma_z $. This minimal two-operator representation appears across several contexts, including counterdiabatic driving~\cite{steveTamingQS}, reinforcement-learning-based quantum control~\cite{zhang2019reinforcement}, exact analytic control constructions for driven two-level systems~\cite{barnes2012analytically}, and transitionless quantum driving~\cite{del2012assisted}. 
 In addition, we include the Hamiltonian formulated as $H = \omega_z\sigma_z + u(t)\omega_x\sigma_x$ in~\cite{kosut2013robust} where the authors develop a framework for robust gate design using a driven single-qubit Hamiltonian. To have a wider representation of single-qubit systems, we also include $H = \sigma_x + u(t)(\sigma_x + \sigma_z)$ from~\cite{lin2020time} which considers time-optimal control of a dissipative qubit under a bounded single control field.  

Two-qubit systems encompass a richer variety of interactions, and the general Hamiltonian takes the form:
\begin{equation}
    H(t) = H_{\text{int}} + H_{\text{local}}^{1}(t) + H_{\text{local}}^{2}(t)
    \label{eq:two_qubit_ham}
\end{equation}
where the interaction term--$H_{int}$, which we limit to a subset of the two-body interactions, may include:
\begin{align}
    H_{\text{int}} &= J_x \sigma_x^{1} \sigma_x^{2} + J_y \sigma_y^{1} \sigma_y^{2} + J_z \sigma_z^{1} \sigma_z^{2}
\end{align}
and local control terms:
\begin{align}
    H_{\text{local}}^{i}(t) &= u_x(t)\sigma_x^{i} + u_y(t)\sigma_y^{i} + u_z(t)\sigma_z^{i}
\end{align}
where $J_{x,y,z}$ denote the coupling strengths along the $x$, $y$, and $z$ directions, $\sigma_{k}^{i}$ represents the Pauli operator acting on qubit $i$ with $k \in \{x,y,z\}$, and $i$ indexes the qubit (either 1 or 2).

Following the approach used for single-qubit systems, we begin our study of two-qubit systems with representative Hamiltonians that have been employed in prior work. One such example is: $H = 0.5(-g\sigma_z^1\sigma_z^2 - h(\sigma_z^1+\sigma_z^2) - h(t)(\sigma_x^1+\sigma_x^2))$ which is considered in~\cite{Xikuns4159802341688z}, and $H = \frac{\Delta}{2}(\sigma_z^1 + \sigma_z^2) + \frac{g}{2} (\sigma_x^1\sigma_x^2 + \sigma_y^1\sigma_y^2) +  \frac{\Omega_1(t)}{2}(\sigma_x^1)+\frac{\Omega_2(t)}{2}(\sigma_x^2)$ as studied in~\cite{Chen2025}.

For three-qubit systems, the dimension of the Hamiltonian space increases substantially. Accordingly, we restrict our analysis to a representative set of systems, listed below:
\begin{itemize}[itemsep=1.5ex]
    \item $H = u_1(t)(\sigma_z^1\sigma_x^2)+u_2(t)(\sigma_x^2\sigma_z^3)+u_3(t)(\sigma_x^2)+u_4(t)(\sigma_y^2)$ as in~\cite{Kim2022} (with minor notation changes), 
    \item $H=J(\sigma_z^1\sigma_z^2 + \sigma_z^1\sigma_z^3 + \sigma_z^2\sigma_z^3) + u_x(t)(\sigma_x^1 + \sigma_x^2 + \sigma_x^3)+ u_y(t)(\sigma_y^1 + \sigma_y^2 + \sigma_y^3)+u_z(t)(\sigma_z^1 + \sigma_z^2 + \sigma_z^3) $ from~\cite{stojanovic2023dicke} with the addition of $u_z(t)(\sigma_z^1 + \sigma_z^2 + \sigma_z^3)$ terms for improved controllability, 
    \item $H=u_{12}(t) (\sigma_x^1\sigma_x^2 + \sigma_y^1\sigma_y^2) + u_{23}(t) (\sigma_x^2\sigma_x^3 + \sigma_y^2\sigma_y^3)$ from~\cite{Xu2016} with some notation adjustments, and 
    \item $H = u_1(t)(\sigma_z^1\sigma_z^2 + \sigma_z^2\sigma_z^3 + \sigma_z^1\sigma_z^3)+u_2(t)(\sigma_x^1 + \sigma_x^2 + \sigma_x^3) + u_3(t) (\sigma_z^1 + \sigma_z^2 + \sigma_z^3)$.
    
\end{itemize}

\subsection{Quantum Control and Robustness Analysis}
\noindent\textbf{Quantum Control}:
Among classical quantum control algorithms, GRAPE~\cite{khaneja2005optimal} has established itself as a state-of-the-art method. It optimizes control pulses by iteratively computing gradients of the fidelity with respect to pulse amplitudes.
While GRAPE achieves high fidelities under ideal conditions, it faces limitations such as sensitivity to model uncertainties and parameter drift, susceptibility to local minima in complex control landscapes, and computational scaling challenges for large systems~\cite{GoerzChristiane2015}. 

RL has gained traction as an alternative to classical control algorithms in recent years. In RL, the quantum state (e.g., the density matrix or relevant observables) serve as the RL agent's “state,” while the control pulse amplitudes \(a_t = \{u_1(t), u_2(t), \dots\}\) play the role of “actions.”  After an action is applied, the system evolves according to quantum dynamics (e.g., via the Schrödinger equation~\eqref{eq:schrodinger} or Lindblad master equation~\eqref{eq:lindblad}), yielding a new state.  A fidelity-based reward \(r_t = F(\rho_{\mathrm{target}}, \rho_t)\) quantifies how close the system is to the desired target state.  The RL agent’s objective is thus to sequentially choose actions so as to maximize cumulative reward over the control trajectory.  

Recent works have demonstrated RL advantages including improved performance over classical methods~\cite{zhang2019reinforcement}, robustness to uncertainties~\cite{niu2019universal}, and scalability to many-body systems~\cite{bukov2018reinforcement}. While these works apply RL to specific Hamiltonians, they do not evaluate performance across diverse Hamiltonian sets; consequently, any change to the system or its parameters necessitates retraining the model.  

Among various RL algorithms, SAC~\cite{haarnoja2018soft, Haarnoja2018SoftAA} stands out for continuous-control tasks such as quantum pulse optimization. SAC is an off-policy, actor–critic algorithm rooted in the maximum entropy RL framework. In contrast to other RL methods that optimize solely for expected return, SAC maximizes a combined objective comprising both expected return (reward) and the policy’s entropy. This approach encourages exploration and prevents premature convergence to suboptimal deterministic policies. By re-using past transitions from a replay buffer, SAC achieves sample efficient learning — a crucial feature when environment interactions (e.g., quantum simulations or experiments) are expensive. Empirical studies on continuous-control benchmarks have shown that SAC consistently outperforms prior on-policy and off-policy methods in terms of both learning stability and asymptotic performance~\cite{haarnoja2018soft, Haarnoja2018SoftAA}. These properties — sample efficiency, stability under hyperparameter variation, and natural support for continuous action spaces — make SAC a compelling candidate, and hence the algorithm we use in this work, for quantum control applications where control pulses must be finely tuned and robustness to noise or modeling uncertainty is required.\\

\noindent\textbf{Robustness analysis}: 
In practical quantum control, a major challenge stems from inherent uncertainties, including decoherence processes and other environment-induced noise that degrade performance; Hamiltonian uncertainties, where the actual values of coupling strengths, detunings, or interaction coefficients deviate from their nominal or estimated values; and control errors, such as fluctuations in pulse amplitudes, or calibration drift in the control hardware~\cite{Koswara_2014, PhysRevA.85.052313, PhysRevLett.132.193801}.

To quantify robustness, methods that explicitly account for uncertainties during control design have been developed. For instance, the Sampling-Based Learning Control (SLC) method builds an “augmented system” by sampling uncertain parameters from a prescribed distribution; a control pulse is then optimized across this ensemble, and finally evaluated over many additional random samples to verify performance under broad uncertainty ranges~\cite{dong2015sampling}. Numerical studies show that this approach can yield control pulses whose fidelity remains high even when uncertainties are significant.

Beyond sampling-based optimization, one classical approach to robustness is to consider the sensitivity of the gate or state-transfer fidelity with respect to small perturbations in system parameters. The work in~\cite{oneil2024robustnessdynamicquantumcontrol} formalizes this in terms of a differential sensitivity bound, which quantifies how much fidelity error may increase under infinitesimal small parameter variations. The aforementioned study introduced a framework in which the upper bound on the differential of the fidelity error with respect to parametric uncertainties gives a guarantee: one can compute a maximum permissible perturbation magnitude that ensures fidelity remains above a target threshold.

Although useful for establishing rigorous worst-case guarantees, differential sensitivity bounds have fundamental limitations such as not being able to clearly indicate the drop in fidelity, and sensitivity score falling when the fidelity is relatively low.

To overcome the limitations of differential bounds and provide a more practical robustness metric, authors in~\cite{PhysRevA.107.032606} have introduced the Robustness-Infidelity Measure ($RIM_p$) as a statistically meaningful, distribution-level measure of robustness and fidelity combined. $RIM_p$, is defined as the p-th order Wasserstein distance between the fidelity distribution (obtained under a specified uncertainty model) and the ideal distribution at unity fidelity. $RIM_1$ — which reduces to the mean infidelity under the uncertainty distribution — is often used due to its simplicity and intuitive interpretation as an average-case robustness metric. We use this metric as the main robustness measure in this work.

%==============================================================================
\section{Methodology and Experimental Setup} \label{sec:methodology_and_experimental_setup}
%==============================================================================
We assume the given system is prepared in some initial pure state $|\psi_0\rangle$, and the goal of the protocol is to reach a given (pure) target state $|\psi_{\text{target}}\rangle$ by optimizing control pulses $\{u_j(t)\}_{t=0}^{T}$ that maximize state transfer fidelity. While the system remains pure under closed unitary evolution, the open-system dynamics governed by Eq.~\eqref{eq:lindblad} generally result in a transition to a mixed state described by a density matrix $\rho(t)$. To evaluate the success of the control under these dissipative conditions, we utilize the fidelity between the target $|\psi_{\text{target}}\rangle$ and the evolved state $\rho(T)$ given by:
\begin{equation}
    F = \langle\psi_{\text{target}}|\rho(T)|\psi_{\text{target}}\rangle
    \label{eq:fidelity}
\end{equation}

Rather than training separate task-specific agents for each Hamiltonian, we formulate a multi-task learning problem where a single policy $\pi_\theta(a|s, \mathbf{c})$ is conditioned on a system descriptor $\mathbf{c}$ that encodes the Hamiltonian structure:
\begin{equation} \label{eq:condition_vector}
\mathbf{c} = [c_1, c_2, c_3, c_4, c_5, c_6]
\end{equation}
where $c_1$, represents the system size, which distinguishes between single-qubit, two-qubit, and three-qubit configurations. The second element, $c_2$, specifies the number of available control parameters or pulses, while $c_3$ defines the strength of the static field within the system. The final three elements, $c_4, c_5$, and $c_6$, serve as binary indicators identifying whether the Hamiltonian includes $\sigma_x$, $\sigma_y$, and $\sigma_z$ Pauli operators, respectively.

\subsection{Open-System SAC Framework}

The quantum control environment, which is based on Gymnasium~\cite{brockman2016openai}, performs initialization such as $\rho_0 = |\psi_0\rangle\langle\psi_0|$, step dynamics (integrate Lindblad master equation~\eqref{eq:lindblad} for time $\Delta t = T/N$, calculate fidelity-based reward signal with optional control regularization), and returns observations such as density matrix elements and derived observables. \\

\noindent\textbf{Reward design:}  
At each time step the agent receives a reward \(r_t\) that depends on the change in fidelity compared to the previous time step, plus additional shaping terms to encourage high-fidelity control. By letting \(F_t = F(\rho_{\mathrm{target}}, \rho_t)\) denote the fidelity at time \(t\), the reward is computed as  
\[
    r_t = 
    \begin{cases}
        100 \cdot (F_t - F_{t-1}), & \text{if } t>0,\\
        10 \cdot F_t, & \text{if } t=0,
    \end{cases}
\]
then augmented by bonuses: if \(F_t > 0.9\) then \(+5\), and if \(F_t\) exceeds a preset threshold \(F_{\min}\) (minimum acceptable fidelity) then \(+20\). Finally, a small time-step penalty \(-0.01 \, t\) is applied to discourage unnecessarily long control sequences.  \\

\noindent\textbf{Adaptive Temporal Optimization:}
In contrast to conventional methods that rely on fixed $T$ and $N$, our approach allows the agent to treat these as learnable parameters. By employing a variable episode length corresponding to different values of $N$, the agent identifies the optimal temporal configuration $T$ during the learning process. The reward signal is designed to penalize longer sequences and evolution times, effectively steering the agent toward faster, more efficient control solutions.\\

\noindent\textbf{Network Architecture:}
The SAC is implemented using the Stable Baselines3~\cite{stable-baselines3} RL framework. The architecture is designed for multi-task learning, utilizing a Policy network $\pi_\theta(a|s, \mathbf{c})$ that maps the combined state and task-specific condition vectors to an action distribution through a Multi-Layer Perceptron (MLP) architecture consisting of five hidden layers $[64, 256, 512, 256, 64]$ with the ReLU~\cite{nair2010rectified} activation function. The input to the policy network is a fixed-length vector formed by concatenating the quantum state representation — density matrix populations and coherences, zero-padded to $d_{\max}^2$ (where $d_{\max}=8$  is the largest Hilbert space dimension for the three-qubit systems) — with the six-dimensional task condition vector $\mathbf{c}$ given in Eq.~\eqref{eq:condition_vector}. The policy network outputs an action vector comprising the evolution time $T$, the number of discretization steps $N$, and the control pulse amplitudes, which are mapped from the policy's native [-1,1] range to physically meaningful values at each environment step.
The agent employs twin Q-networks $Q_{\phi_1}$ and $Q_{\phi_2}$ with an identical MLP structure to the policy network to estimate the action-value functions. The entropy parameter $\alpha$ is dynamically adjusted during training to maintain a balance between exploration and exploitation.

\subsection{Training Details}
\noindent\textbf{Closed System Training:}
We train both task-specific and multi-task control models under closed-system dynamics. A fidelity threshold of $0.999$ serves as the episode termination criterion, after which the environment resets, and a new training episode begins. These models serve as a baseline for comparing task-specific and multi-task learning under ideal conditions.\\

\noindent\textbf{Open System Training:} % with Progressive Hamiltonian Expansion:} 
To account for realistic decoherence effects, we incorporate Lindblad dissipators given by Eq.~\eqref{eq:lindblad} directly into the SAC training environment and train several multi-task models under open-system dynamics. Building on the parity established between multi-task and bespoke models in our closed-system benchmarks—where the multi-task framework demonstrated competitive performance—our open-system analysis focuses exclusively on the multi-task architecture. This approach maintains computational tractability, while allowing us to focus on the primary objective of developing a multi-task agent. After models are trained, we perform a comprehensive analysis of each model's performance, including comparisons across configurations (e.g., adaptive $T$ and $N$ versus fixed $T$ and $N$) and evaluation under varying decoherence rates.\\

\noindent\textbf{Progressive Hamiltonian training set expansion:}
To further investigate the degree of training-set diversity required for effective generalization in open-system settings, we adopt a progressive Hamiltonian expansion strategy. A base set of five representative Hamiltonians from literature (whose details are presented in Table~\ref{tab:Initial-Hamiltonian-details}) spanning single-qubit, two-qubit, and three-qubit systems is used to initialize the training set. A multi-task model is trained on this initial set and evaluated on the full and held-out Hamiltonian sets. The initial and target states are either drawn from the corresponding literature where available, or chosen to represent a diverse range of state preparation tasks.
 
\begin{table}[ht]
    \centering
    \caption{Initial Hamiltonian set: single-qubit, two-qubit, and three-qubit systems with their corresponding initial and target states. These Hamiltonians are taken from the corresponding literature. The goal of the time evolution operation is to transfer population from the initial states to the target states.} 
    \label{tab:Initial-Hamiltonian-details}

    \begin{tabular}{p{6.0cm}C{2.5cm}C{2.5cm}}
    
        \toprule
        System (Hamiltonian) & Initial State & Target State\\
        \toprule
        
        $H = \omega_z\sigma_z + u(t)\omega_x\sigma_x$; $\omega_{x,z}=1$~\cite{kosut2013robust} & $\ket{0}$ & $\ket{1}$  \\
        \midrule
        $H = \sigma_x + u(t)(\sigma_x + \sigma_z)$~\cite{lin2020time} & $(\ket{0}+\ket{1})/\sqrt(2)$ & $\ket{1}$  \\
        \midrule
        $H = \frac{1}{2}(-j\sigma_z^1\sigma_z^2 - h(\sigma_z^1+\sigma_z^2) - u(t)(\sigma_x^1+\sigma_x^2))$; $j=h=1 $~\cite{Xikuns4159802341688z} & $(\ket{00}+\ket{01}+\ket{10}+\ket{11})/\sqrt(4)$ & $\ket{00}$ \\
        \midrule
        $ H = \frac{1}{2}(\sigma_z^1 + \sigma_z^2 + j(\sigma_x^1\sigma_x^2 + \sigma_y^1\sigma_y^2) +  {u_1(t)}\sigma_x^1+u_2(t)\sigma_x^2)$; $j=1$~\cite{Chen2025}  & $\ket{00}$ & $(\ket{01}+\ket{10})/\sqrt(2)$  \\
        \midrule
        $H = u_1(t)(\sigma_z^1\sigma_x^2)+u_2(t)(\sigma_x^2\sigma_z^3)+u_3(t)(\sigma_x^2)+u_4(t)(\sigma_y^2)$~\cite{Kim2022} & $\ket{000}$ & $-i\ket{010}$ \\
        \bottomrule
    \end{tabular}
\end{table}
The training set is then iteratively enlarged by sampling additional Hamiltonians from the complete Hamiltonian space generated using Eq.~\eqref{eq:single_qubit_ham} for single-qubit systems, Eq.~\eqref{eq:two_qubit_ham} for two-qubit systems, and the three-qubit systems described in Section~\ref{hamiltonian_details}. Successive expansions increase the size of the training set from 5 to 10, 20, 30, 40, ultimately reaching the full collection of 51 Hamiltonians. At each stage, the model is retrained and its performance is reassessed on both the unseen and entire Hamiltonians. This procedure enables us to evaluate how well the model learns shared quantum control principles across different Hamiltonians for practical generalization. Details of these experiments and results is presented in Section~\ref{multi_task_learning_results}.

\subsection{Robustness Infidelity Measure (RIM)}\label{subsec:RIM}
To quantify control robustness, we use RIM introduced in~\cite{PhysRevA.107.032606}:
\begin{equation}
\text{RIM} = \frac{1}{M}\sum_{m=1}^{M} (1 - F_m(\delta_m))
\label{eq:rim}
\end{equation}
where $F_m$ is the fidelity under perturbation $\delta_m$ drawn from specified distributions. We evaluate RIM under the following three perturbation scenarios:\\

\noindent\textbf{Pulse amplitude perturbations:}
After control pulses that result in successful population transfer are obtained using either SAC or GRAPE, we apply perturbations to these control pulses given by:
\begin{equation}
u_j^{\text{p}}(t) = u_j(t) + \epsilon_j
\end{equation}
where $u_j^{\text{p}}(t)$ is the perturbed pulse amplitude, $\epsilon_j \sim \mathcal{U}(-\delta_u, \delta_u)$ with $\delta_u$ controlling perturbation magnitude. We set $\delta_u = 0.05$ in our experiments, (corresponding to $5\%$ of the maximum pulse amplitude), a value chosen to reflect typical experimental uncertainties in control hardware~\cite{niu2019universal}. \\
%which we select it to be 5\% of the maximum amplitude of the pulse

\noindent\textbf{Decoherence Rate Perturbations:}
Under perturbation, the decoherence rate undergoes a shift from its nominal value $\gamma_k$ such that:
\begin{equation}
\gamma_k^{\text{p}} = \gamma_k + \epsilon_k
\end{equation}
where $\gamma_k^{\text{p}}$ is the perturbed decoherence rate, $\gamma_k$ is the nominal decoherence rate before perturbation which is set to $0.01$, and $\epsilon_k \sim \mathcal{U}(0, \delta_\gamma)$, $\delta_\gamma$ representing the perturbation amount. We set $\delta_\gamma = 0.005$ (50\% of the max decoherence rate--0.01), this range is chosen to reflect the significant temporal fluctuations in decoherence times observed in experimental platforms~\cite{dasgupta2023adaptivemitigationtimevaryingquantum, PhysRevLett_121_090502}. Any value less than this was found to be too small to make a noticeable difference. \\

\noindent\textbf{Combined Perturbations:}
Both pulse amplitude and decoherence rate perturbations applied simultaneously.

After generating the perturbations as above, they are applied to the GRAPE based and SAC based control results, for comparison.

\subsection{Simulation Parameters:}
\noindent\textbf{Control Constraints:} The following control constraints are placed on the agent: 
\begin{itemize}
    \item Maximum control amplitude: $|u_j(t)| \leq u_{\max}$ = 1, with additional scaling factor applied based on system requirements (for example some Hamiltonians work better when the control amplitude is higher than the static field control strength). 
    \item Evolution time range: $T \in [1, 20]$ 
    \item Number of segments: $N \in [2, 60]$, we limit $N_{max} \leq 3T_{max}$, where $T_{max}$ is the max evolution time allowed (20 time units), thereby preventing overly fine segmentation without rendering the control discretization too coarse for successful system evolution.
\end{itemize} 

\noindent\textbf{SAC Hyper-parameters:}
The hyper-parameters in Table~\ref{tab:sac_params} are used in all SAC models.
\begin{table}[ht]
    \centering
    \caption{SAC Training Hyperparameters}
    \label{tab:sac_params}

    \begin{tabular}{p{6.5cm}C{3.0cm}}
        \toprule
        Parameter & Value \\
        \midrule
        Learning rate (actor, critic) & 3*10e-4 \\
        \midrule
        Discount factor & 0.98 \\
        \midrule
        Soft update coefficient  & 0.05 \\
        \midrule
        Replay buffer size & 500,000 \\
        \midrule
        Batch size & 64 \\
        \midrule
        Initial temperature(entropy) & auto \\
        \midrule
        Hidden layer sizes & [64, 256, 512, 256, 64] \\
        \bottomrule
    \end{tabular}
\end{table}

%==============================================================================
\FloatBarrier
\section{Results and Discussion} \label{sec:results_and_discusion}
%==============================================================================
This section presents a comprehensive analysis of the experimental results and explores their broader implications. Performance is quantified primarily through state fidelity and RIM. To ensure statistical significance, we perform 30 state transfer experiments per Hamiltonian. Within each experiment, reported performance represents the maximum fidelity attained across 25 inference trials. Details of each Hamiltonian, initial states and target states used in our experiment is presented in Appendix~\ref{appendix:A} (Tables~\ref{tab:single-qubit-hamiltonians}, \ref{tab:two-qubit-hamiltonians}, and \ref{tab:three-qubit-hamiltonians}).

To establish a baseline and assess the effectiveness of the multi-task SAC agent in the absence of decoherence, we train all systems under closed-system (unitary) dynamics. This includes task-specific models trained individually for each Hamiltonian, as well as a single multi-task model trained on all Hamiltonians. This allows us to quantify the maximum fidelities achievable in the task-specific setting and to evaluate how performance changes when transitioning to a unified multi-task model.

\begin{figure}[htbp]
    \centering
    \includegraphics[width=0.99\textwidth]{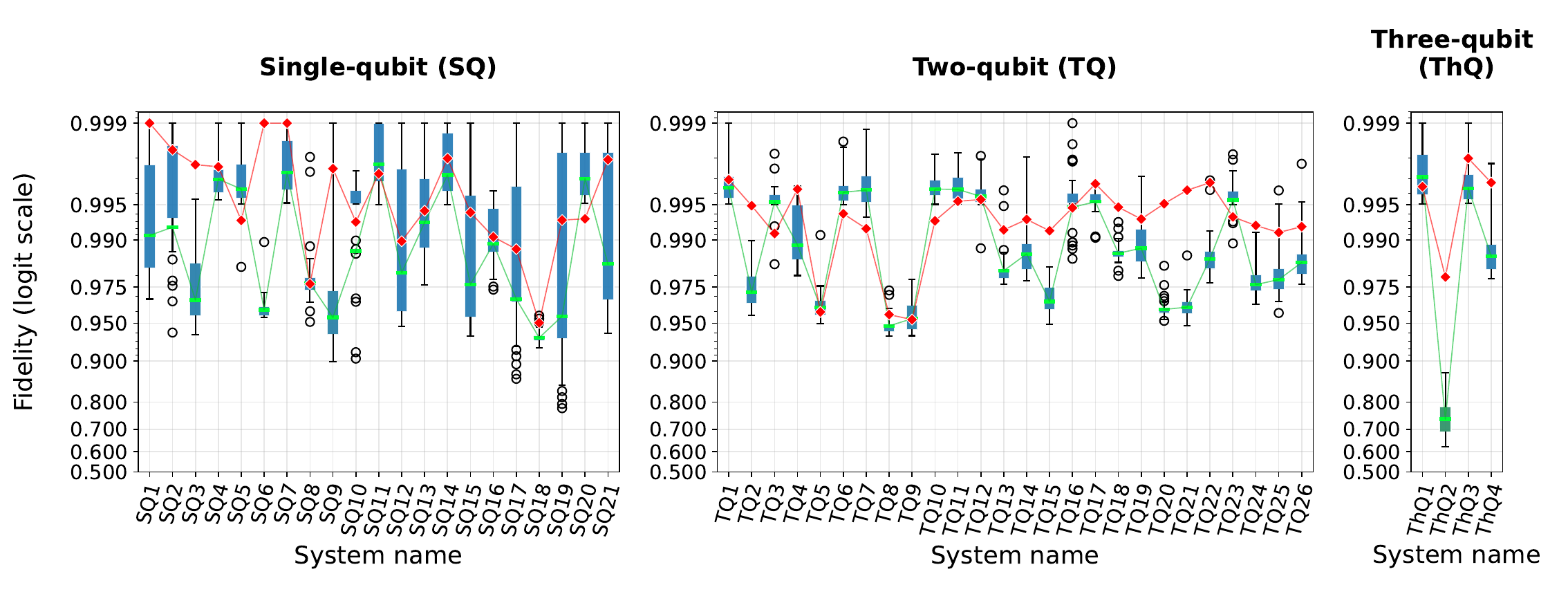}
    \caption{Distribution of maximum state fidelities for closed-system dynamics across single-qubit (SQ), two-qubit (TQ), and three-qubit (ThQ) systems. Each box plot represents 30 independent experiments, demonstrating the agent's capability to reach near-ideal fidelities across a diverse Hamiltonian set. The red diamonds denote the average fidelity for task-specific models, while the box plot represents fidelity distribution for multi-task model(the horizontal green lines denote the mean fidelity).}
    \label{fig:closed_system_results}
\end{figure}

Figure~\ref{fig:closed_system_results} summarizes these results: red diamonds denote the average closed-system fidelity achieved by task-specific SAC models for each Hamiltonian, while the box plots illustrate the fidelity distribution obtained by the multi-task model. As shown in the figure, the difference between the mean fidelities for the task-specific models (red diamonds) and the multi-task model (green horizontal lines) is less than $0.005$ for most systems. In some cases, the multi-task model even outperforms the task-specific ones, offering significant practical advantages as a single, unified controller. Notably, the multi-task agent achieves fidelities as high as $0.999$ for several systems, demonstrating the ability of the multi-task SAC agent to discover near-optimal control sequences under idealized conditions. 

After confirming the agent's baseline capability, we trained a separate model under open-system dynamics. Sections~\ref{multi_task_learning_results},~\ref{subsec:robustness_analysis},~\ref{subsec:adaptive_optimization}, and~\ref{subsec:closed_system_open_system_performance} present these experimental results, with Section~\ref{subsec:closed_system_open_system_performance} specifically comparing closed-system and open-system performance for selected Hamiltonians. For this open-system model, optimization is terminated once the state fidelity reaches a predefined minimum fidelity threshold of $0.95$. While this value is below the requirements for practical quantum applications, it was selected to account for the inherent limitations of certain open-system dynamics. By adopting this threshold, we ensure that the performance evaluation focuses on the SAC agent's learning capabilities rather than the physical constraints of certain systems that struggle to reach higher fidelities under noise.

To calculate the RIM scores, we generate a comprehensive dataset involving $51$ Hamiltonians. For each system, we record control pulses across 30 experiments, resulting in 1,530 data points each for the multi-task SAC and GRAPE approaches. Data points with pre-perturbation fidelity below the predefined threshold of $0.95$ are excluded from the RIM analysis. This filtering ensures that the robustness evaluation is not biased by systems that cannot achieve the target fidelity, particularly certain three-qubit systems, where the limitation arises from control performance rather than robustness to uncertainty. As mentioned in Section~\ref{subsec:RIM}, the RIM is calculated by subjecting the nominal control pulses to two primary noise sources: pulse amplitude fluctuations, and decoherence rate variations, as well as their combined effects. For each control sequence, we generate $n$ noisy samples (we used $n=15$) and compute the resulting fidelities through open-system simulations. The RIM, for each Hamiltonian, is then defined as the average infidelity across these samples, providing a quantitative metric for how sensitive a specific control solution is to experimental uncertainties and environmental noise. This process is repeated across all Hamiltonians to derive both system-specific and aggregate robustness statistics.

We run similar experiments using pulses generated by SAC and GRAPE. Our goal is to identify how systems respond to uncertainties on control pulses generated using the two approaches. 

\subsection{Multi-Task Learning under Open System Dynamics} \label{multi_task_learning_results} %and Hamiltonian Generalization
\begin{figure}[htbp]
    \centering
    \includegraphics[width=0.99\textwidth]{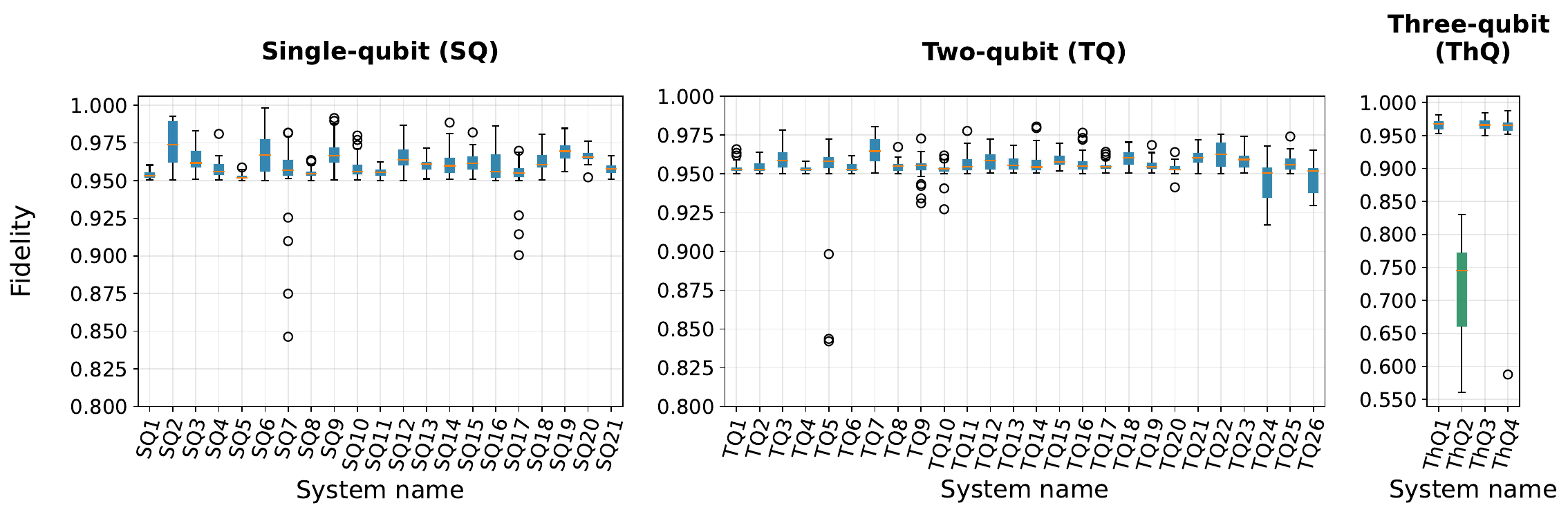}
    \caption{Open-system fidelity distribution across various quantum control systems. The box plots illustrate the distribution of fidelities for single-qubit ($SQ$), two-qubit ($TQ$), and three-qubit ($ThQ$) systems over 30 experiments per system. The horizontal orange lines denote the median fidelity.}
    \label{fig:fidelity_boxplot}
\end{figure}

\noindent\textbf{Fidelity distribution across systems:}
Figure~\ref{fig:fidelity_boxplot} presents the open-system fidelity distribution for the multi-task SAC model across all systems (we refer the reader to Appendix~\ref{appendix:A} for system details). 

As illustrated in Figure~\ref{fig:fidelity_boxplot}, almost all evaluated systems achieve a median fidelity exceeding 0.95, demonstrating the effectiveness of the SAC-based approach across varying system sizes. All single-qubit systems ($SQ^*$) exhibit median fidelities above 0.95. While two-qubit systems ($TQ^*$) face increased Hilbert space complexity, they maintain higher performance, with very few systems having fidelities below the 0.95 threshold suggesting that while the agent generally finds near-optimal solutions, it occasionally encounters ``stiff'' trajectories under specific stochastic conditions. 

Three-qubit systems ($ThQ^*$) generally exhibit high fidelities comparable to lower-dimensional systems.
A notable exception is the three-qubit all-to-all Ising-coupled model utilizing global $X$ and $Y$ controls (ThQ2, see system details in Appendix~\ref{appendix:A}). This specific configuration, tasked with a population transfer from the initial state $\ket{000}$ to the Dicke target state $\ket{D_3^1} = \frac{1}{\sqrt{3}}(\ket{100} + \ket{010} + \ket{001})$, presents significant control challenges due to increased complexity. Nevertheless, the model successfully achieves fidelities above 0.95 for the remaining three-qubit configurations. These results suggest that SAC-based models serve as a robust alternative to traditional optimal control techniques, provided they are trained on a sufficiently diverse distribution of system samples and state transformations.\\

\noindent\textbf{Progressive Hamiltonian Expansion Results:} 
To evaluate the generalization capabilities of our framework, we progressively increase the diversity of the training set, starting with five representative Hamiltonians. This approach allows us to assess how effectively the model captures the underlying physical relationships across different systems—for instance, determining whether an agent trained on a specific subset of single-qubit Hamiltonians can perform successful state transfer on unseen systems within the same family. Initially, the model was trained using only the Hamiltonians detailed in Table~\ref{tab:Initial-Hamiltonian-details}. Despite this limited exposure, the resulting policy achieved a successful evolution with success rate of approximately 37\% when evaluated across the full suite of 51 Hamiltonians which includes successful control of systems entirely absent from the training set, suggesting a degree of zero-shot transferability inherent in the learned representation. The experiments were performed on 5, 10, 20, 30, and 40 Hamiltonians, and the results are summarized in Table~\ref{tab:progressive-hamiltonian-expansion}.

\begin{table}[ht]
    \centering
    
    \caption{Progressive expansion of the multi-task training set. As the model is exposed to more Hamiltonians, full-set performance improves significantly. Held-out success rates are influenced by the decreasing availability of simpler single-qubit systems in the test pool as the training set expands. Note, we define success rate as percentage of experiments that resulted in a fidelity of 0.95 or above.} 
    \label{tab:progressive-hamiltonian-expansion}
    
    \begin{tabular}{C{3.5cm}C{3.5cm}C{3.5cm}}
    
        \toprule
        Number of Hamiltonians the model is trained on & Inference success rate(\%) on full Hamiltonian set & Inference success rate(\%) on held-out Hamiltonian set\\
        \toprule
        
        5 Hamiltonians & 37.3 & 30.5 (14/46 systems)\\
        \midrule
        10 Hamiltonians & 38.2 & 23.6 (9/41 systems)\\
        \midrule
        20 Hamiltonians & 52.7 & 25.2 (6/31 systems)\\
        \midrule
        30 Hamiltonians & 68.5 & 32.2 (7/21 systems)\\
        \midrule
        40 Hamiltonians & 77.7 & 37.0 (3/11 systems)\\
        \midrule
        51 Hamiltonians & 95.2 & -\\
        \bottomrule
    \end{tabular}
\end{table}

The agent's performance on held-out samples demonstrates that it can reliably solve unseen single-qubit systems with fidelities exceeding 0.95. As shown in Table~\ref{tab:progressive-hamiltonian-expansion}, the model trained on only five Hamiltonians achieved successful state preparation (fidelity $\geq 0.95$) on 14 out of 46 unseen systems. This performance on diverse initial and target state requirements suggests that the agent did not merely memorize specific training instances but successfully extracted shared quantum control principles across systems. 

The initial success rate on held-out samples—30.5\% with only five training Hamiltonians—is primarily attributed to the agent's high proficiency with single-qubit (SQ) systems. However, as the training set expands, these ``easier'' single-qubit systems are progressively transitioned from the held-out set into the training set. Consequently, the marginal increase in success rate diminishes, as the remaining held-out set becomes increasingly dominated by more complex, multi-qubit interactions. Despite this, the agent demonstrates ability to drive the population toward the target states across diverse two-qubit and three-qubit Hamiltonians with fidelities not far from the 0.95 threshold, as reported in Appendix~\ref{appendix:B}.

\subsection{Robustness Analysis: SAC vs. GRAPE}\label{subsec:robustness_analysis}
\begin{figure}[hbp]
    \centering
    \includegraphics[width=0.95\textwidth]{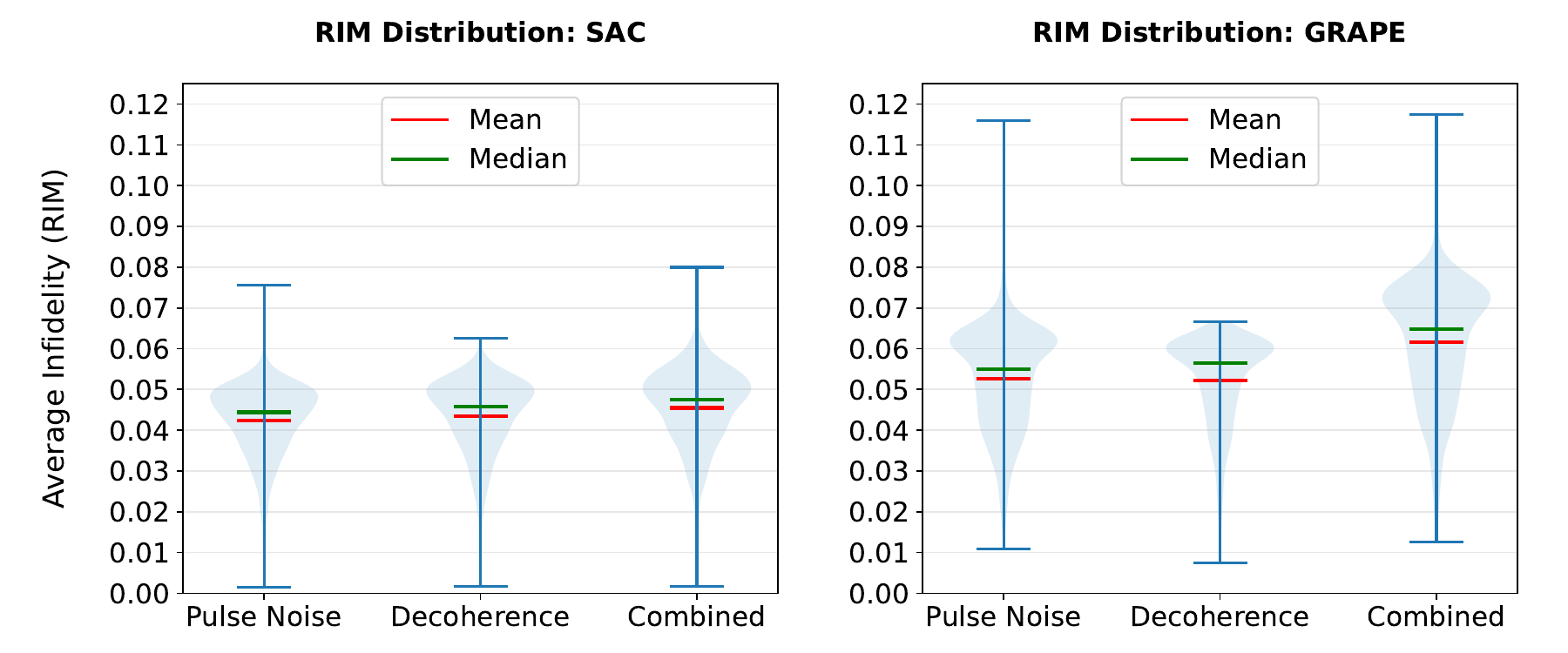}
    \caption{Robustness Infidelity Measure (RIM) analysis for SAC and GRAPE. Left: RIM for SAC, Right: RIM for GRAPE. SAC achieves mean infidelity of approximately 0.045 under all perturbation types, and GRAPE exhibits higher mean infidelity of approximately 0.061 under combined perturbations.}
    \label{fig:sac_grape_rim}
\end{figure}

The RIM for the SAC and GRAPE control strategies is presented in Figure~\ref{fig:sac_grape_rim}. From this figure we can observe that, SAC achieves a consistently lower median/mean RIM scores across all perturbation types compared to GRAPE. 
While both methods show slightly increased sensitivity under combined perturbations, GRAPE exhibits a more significant degradation in performance, with its mean average infidelity rising to approximately 0.061. In contrast, SAC maintains a notably tighter distribution and a lower mean infidelity of roughly 0.045, even when subjected to multiple simultaneous noise sources. It is also worth remembering that GRAPE operates on a per system level, whereas SAC is seeking to solve all systems simultaneously. 

Several factors contribute to the improved performance of SAC over GRAPE. First, the stochastic nature of SAC policies ensures the sampling of diverse control trajectories, effectively covering a wider range of operating conditions during training. Second, the maximum entropy objective prevents premature convergence to suboptimal solutions. Furthermore, training under Lindblad dynamics explicitly accounts for open-system effects, yielding controls that are inherently less sensitive to decay rate variations. Finally, whereas GRAPE relies on precise gradient computations that are susceptible to model inaccuracies, SAC leverages a replay buffer of diverse experiences to learn a robust control distribution rather than a singular, rigid path. This inherent robustness to various perturbation types makes SAC-based control a more reliable alternative for sensitive quantum systems that demand minimal performance fluctuation and high resilience against experimental uncertainties.

\subsection{Adaptive Temporal Optimization} \label{subsec:adaptive_optimization}

\begin{figure}[htbp]
    \centering
    \includegraphics[width=0.95\textwidth]{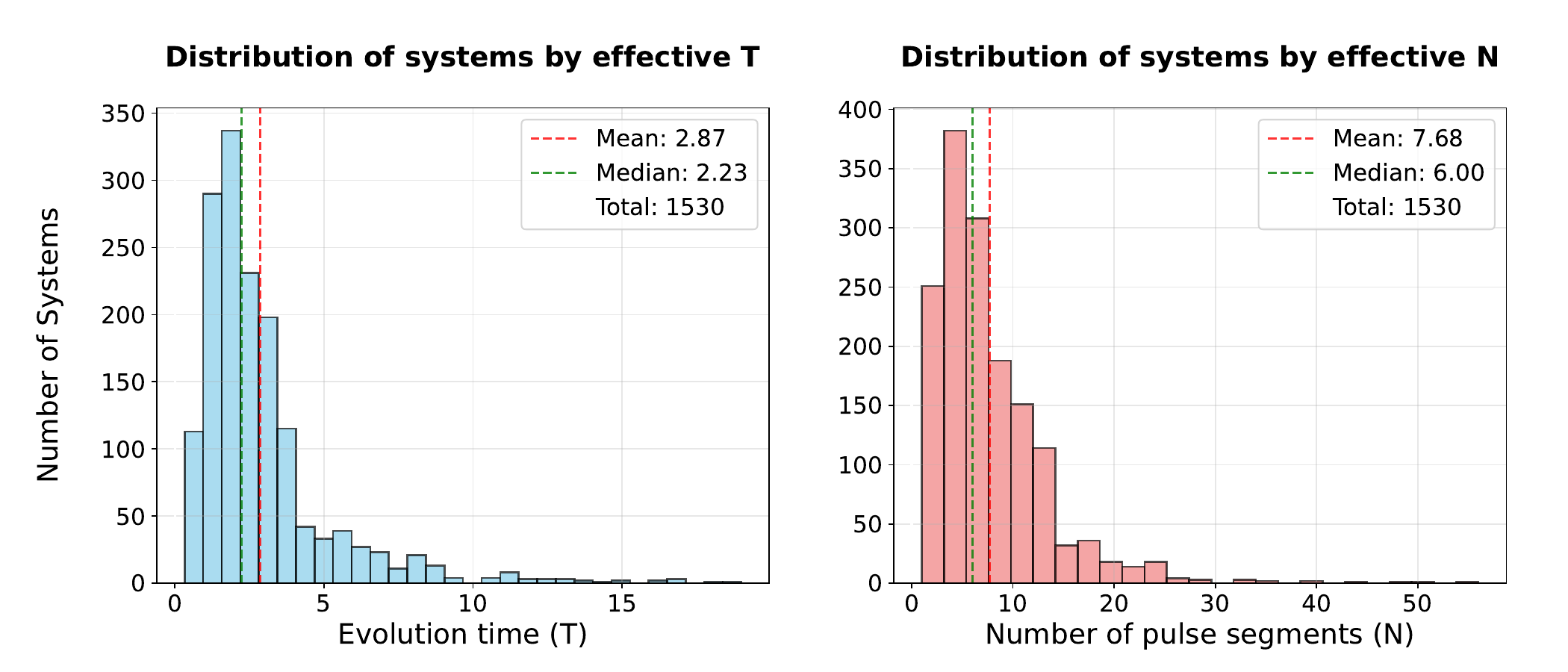}
    \caption{Distribution of learned effective evolution time $T$ (left) and effective number of control steps $N$ (right) across 1530 experiments. The distributions show that most systems converge to relatively short evolution times (median $T = 2.23$, mean $T = 2.87$) and moderate discretization (median $N = 6$, mean $N = 7.68$). Note that, $N$ assumes an integer value.}
    \label{fig:temporal_distribution}
\end{figure}

As demonstrated by the results in Table~\ref{tab:progressive-hamiltonian-expansion}, the SAC agent successfully learns to optimize both the evolution time $T$ and the number of segments $N$ simultaneously. With adaptive $T$ and $N$, the success rate of our experiments was 95.2\%. In contrast, performance degrades when these parameters are held constant. A model trained with fixed values, specifically $T=10$ and $N=20$, which are chosen experimentally to reflect values found individually in prior works~\cite{xiao20222, zhang2019reinforcement}, resulted in a lower success rate of 93.5\%. This shows having an adaptive $T$ and $N$, has advantages such as: improved success rate because the model finds the best values of $T$ and $N$ that will result in improved fidelity, reduced average evolution time, $T$,  as can be seen in Figure~\ref{fig:temporal_distribution} that helps in minimizing decoherence accumulation, and computational efficiency due to reduced average values of $N$.

Figure~\ref{fig:temporal_distribution} shows the distribution of learned temporal parameters across all systems. From this figure, we can see that for most systems, the evolution time is $T < 5$ time units, number of segments $N$ has a median $N = 6$ which suggests that many control problems can be solved with relatively short evolution time and coarse pulse sequences, reducing experimental complexity. 

\subsection{Closed-System vs. Open-System Performance} \label{subsec:closed_system_open_system_performance}
In this specific experiment, we investigate how a model trained under idealized closed-system conditions responds when subjected to open-system dynamics. This is important to reveal the extent to which decoherence degrades control performance and to motivate the need for models trained directly under open-system conditions.\\

\noindent\textbf{Closed-System Baseline Results:}
We trained the five Hamiltonians given in Table~\ref{tab:Initial-Hamiltonian-details} in a task-specific manner using noise-free closed-system evolution. To assess the models' robustness to environmental coupling, we performed inference under open-system dynamics governed by Eq.~\eqref{eq:lindblad}, introducing decoherence rates of $\gamma = 0.01$ and $\gamma = 0.02$. This evaluation, summarized in Table~\ref{tab:closed-open-system-results}, highlights the performance degradation inherent in models lacking exposure to dissipative effects during the training phase. 

\begin{table}[ht]
    \centering
    \caption{Fidelity and success rate comparison of SAC agent performance across different Hamiltonian systems, a model trained with closed-system dynamics applied to open-systems. Note: $^*$ corresponds to $\gamma=0.01$, and $^{**}$ correspond to $\gamma=0.02$, success rate is defined as percentage of experiments that resulted in a fidelity of 0.95 or above.}
    \label{tab:closed-open-system-results}
    \begin{tabular}{L{4.5cm}C{0.70cm}C{1.10cm}C{0.70cm}C{1.10cm}C{0.70cm}C{1.10cm}}
        \toprule
        \multirow{2}{*}{} & \multicolumn{2}{c}{\textbf{Closed System}} & \multicolumn{2}{c}{\textbf{Open System$^*$}} & \multicolumn{2}{c}{\textbf{Open System$^{**}$}}\\
        \cmidrule(lr){2-3} \cmidrule(lr){4-5} \cmidrule(lr){6-7}
        \textbf{System Hamiltonian} &  Mean Fidelity & Success rate (\%) &  Mean Fidelity & Success rate (\%) &  Mean Fidelity & Success rate (\%)\\
        \toprule
        $H = \omega_z\sigma_z + u(t)\omega_x\sigma_x$; $\omega_{x,z}=1$ & 0.9564 &  100 & 0.9513 & 93.3 & 0.9278 & 0\\
        \midrule
        $H = \sigma_x + u(t)(\sigma_x + \sigma_z)$ & 0.9624 & 100 & 0.9529 & 100 & 0.9244 & 13.3\\
        \midrule
        $H = \frac{1}{2}(-j\sigma_z^1\sigma_z^2 - h(\sigma_z^1+\sigma_z^2) - u(t)(\sigma_x^1+\sigma_x^2))$; $j=h=1 $  &  0.9541 & 100 & 0.9509 & 90 & 0.9236 & 0\\
        \midrule
        $ H = \frac{1}{2}(\sigma_z^1 + \sigma_z^2 + j(\sigma_x^1\sigma_x^2 + \sigma_y^1\sigma_y^2) +  {u_1(t)}\sigma_x^1+u_2(t)\sigma_x^2)$; $j=1$ & 0.9583 & 100 & 0.9333 & 13.3 & 0.9108 & 0\\
        \midrule
        $H = u_1(t)(\sigma_z^1\sigma_x^2)+u_2(t)(\sigma_x^2\sigma_z^3)+u_3(t)(\sigma_x^2)+u_4(t)(\sigma_y^2) $ & 0.9576 & 100 & 0.9687 & 100 & 0.9707 & 100\\
        \bottomrule
    \end{tabular}
\end{table}

While the closed-system model maintains the ability to drive certain systems to the target state under moderate decoherence, increased noise levels significantly hinder the agent's capacity to achieve high-fidelity state transfer. This performance degradation in Table~\ref{tab:closed-open-system-results} demonstrates that closed-system training can be insufficient for practical quantum control, as policies optimizing for high fidelity in isolated models may not generalize to environments where irreversible information loss occurs due to environmental coupling.\\

\noindent\textbf{Open-System Retraining Results:} 
To evaluate the efficacy of noise-aware training, we retrained the SAC agent using the same five Hamiltonians under explicit Lindblad dynamics, resulting in the substantial performance gains documented in Table~\ref{tab:open-system-results}. While the training phase utilized a fixed decoherence rate of $\gamma=0.01$, inference was conducted at both $\gamma=0.01$ and $\gamma=0.02$. This approach allows us to investigate the model's generalization capabilities when subjected to environmental noise levels that exceed the training distribution. 

\begin{table}[ht]
    \centering
    \caption{Open-system evaluation results: The model is trained using open-system dynamics with decoherence rate of $\gamma=0.01$ and testing is also performed using open-system (Lindblad) dynamics with $\gamma=0.01$ and $\gamma=0.02$. Note: $^*$ corresponds to $\gamma=0.01$, and $^{**}$ corresponds to $\gamma=0.02$, success rate is defined as percentage of experiments that resulted in a fidelity of 0.95 or above.}
    \label{tab:open-system-results}
    \begin{tabular}{L{6.0cm}C{1.0cm}C{1.0cm}C{1.0cm}C{1.0cm}}
        \toprule
        \multirow{2}{*}{} & \multicolumn{2}{c}{\textbf{Open System$^*$}} & \multicolumn{2}{c}{\textbf{Open System$^{**}$}}\\
        \cmidrule(lr){2-3} \cmidrule(lr){4-5}
        \textbf{System Hamiltonian} & Mean Fidelity & Success rate (\%) & Mean Fidelity & Success rate (\%)\\
        \midrule
        $H = \omega_z\sigma_z + u(t)\omega_x\sigma_x$; $\omega_{x,z}=1$ & 0.9601 & 100 & 0.9529 & 96.7\\
        \midrule
        $H = \sigma_x + u(t)(\sigma_x + \sigma_z)$ & 0.9632 & 100 & 0.9597 & 100\\
        \midrule
        $H = \frac{1}{2}(-j\sigma_z^1\sigma_z^2 - h(\sigma_z^1+\sigma_z^2) - u(t)(\sigma_x^1+\sigma_x^2))$; $j=h=1 $ & 0.9595 & 100 & 0.9527 & 100 \\
        \midrule
        $ H = \frac{1}{2}(\sigma_z^1 + \sigma_z^2 + j(\sigma_x^1\sigma_x^2 + \sigma_y^1\sigma_y^2) +  {u_1(t)}\sigma_x^1+u_2(t)\sigma_x^2)$; $j=1$  & 0.9588 & 100 & 0.9521 & 100\\
        \midrule
        $H = u_1(t)(\sigma_z^1\sigma_x^2)+u_2(t)(\sigma_x^2\sigma_z^3)+u_3(t)(\sigma_x^2)+u_4(t)(\sigma_y^2)$ & 0.9572 &  100 & 0.9593 & 100\\
        \bottomrule
    \end{tabular}
\end{table}
The results in Table~\ref{tab:open-system-results} demonstrate that incorporating decoherence into the training loop naturally yields control policies with inherent robustness to perturbations. Notably, the agent maintains high fidelity even when encountering dissipative effects beyond those used in its training phase, suggesting that the maximum entropy objective encourages the discovery of stochastic-stable solutions rather than brittle, noise-specific optimizations.

%==============================================================================
\FloatBarrier
\section{Conclusion} \label{sec:conclusion}

We have presented a multi-task reinforcement learning (RL) framework based on Soft Actor-Critic (SAC) for the control of multiple open quantum system variations. Our results demonstrate that a single multi-task SAC agent effectively performs state transfer across a large number of systems, eliminating the need to train separate agents for individual systems. The framework was validated using Hamiltonians studied in the quantum control literature or those derived using generalized single-qubit and two-qubit Hamiltonians, demonstrating that a single policy can handle diverse and practically relevant systems. This also points toward the broader possibility of unified control models capable of executing large sets of quantum operations, including gate sets across diverse hardware platforms, within a single learned policy. 

Under closed-system settings, a comparison of task-specific and multi-task models revealed that the multi-task framework not only matches but occasionally surpasses specialized models, establishing it as a robust, unified paradigm for high-fidelity system control.

Through progressive Hamiltonian set expansion, we demonstrate that the multi-task agent learns fundamental quantum control principles rather than memorizing specific training configurations. This is evidenced by the agent's ability to successfully perform state transfer tasks on unseen Hamiltonians and initial-target state pairs within the same Hamiltonian space, confirming a robust ability to generalize. 

The robustness analysis reveals that SAC-generated pulses consistently outperform GRAPE-based methods under perturbations in control amplitudes, decoherence rates, and combined noise sources. Beyond improved fidelity and robustness, our framework introduces a technique to generate system specific adaptive evolution time ($T$) and pulse segment count ($N$) by treating these as trainable parameters. This adaptability leads to higher success rates and significantly reduced evolution times, which minimizes the system's susceptibility to decoherence and enhances computational efficiency when compared to fixed-$T$ and fixed-$N$ models. Such robust control strategies that maintain performance under realistic noise profiles may complement quantum error correction schemes by reducing logical error accumulation through noise-resilient control sequences.

Despite these promising results, several limitations remain. The current implementation is restricted to few-qubit systems due to the exponential growth of the Hilbert space, which correspondingly enlarges the agent’s observation space and increases training complexity. Furthermore, the present framework assumes fixed initial and target states for specific one-qubit, two-qubit, and three-qubit systems, limiting its applicability to more general state-transfer scenarios. Consequently, a rigorous controllability analysis has yet to be performed to determine the reachability of arbitrary initial-target pairs under open-system dynamics.

Future work will focus on a scalable architecture capable of handling arbitrary state pairs across variable qubit configurations, moving toward a genuinely universal quantum control policy. Exploring alternative pulse parametrizations—such as Fourier or sinc bases—where the agent outputs basis coefficients rather than raw pulse amplitudes could reduce action space dimensionality and potentially improve convergence. Additionally, experimental deployment on hardware platforms will be crucial for validating real-time adaptability and robustness.

\bigskip
\backmatter

\bmhead*{Data availability:} 

A companion GitHub repository will be available containing source code and sample data generated during this study.

\bmhead*{Acknowledgments:}

This work was funded by Taighde Éireann – Research Ireland  through the Research Ireland Centre for Research Training in Machine Learning (18/CRT/6183). The authors acknowledge the use of the UCD Sonic High Performance Computing (HPC) cluster, provided by UCD IT Services, for the computational work in this research.
%\section*{Declarations}

\begin{appendices}

\section{Complete Hamiltonian Collection} \label{appendix:A}

This appendix presents the complete collection of Hamiltonians used in our experiments. The collection spans single-qubit (21 systems), two-qubit (26 systems), and three-qubit (4 systems) architectures with diverse drift terms, interaction types, and control configurations. Some Hamiltonians are taken directly from papers, others are sampled from the standard forms using Eq.~\eqref{eq:single_qubit_ham} and Eq.~\eqref{eq:two_qubit_ham}.

\begin{longtable}{cL{8.250cm}C{1.0cm}C{1.0cm}}
\caption{Details of single-qubit Hamiltonians used in Multi-Task RL Control} \label{tab:single-qubit-hamiltonians} \\

\toprule
\textbf{ID} & \textbf{Hamiltonian } & \textbf{Initial State} & \textbf{Target State} \\%& \textbf{System Name} \\
\toprule
\endfirsthead

\multicolumn{4}{c}%
{{\tablename\ \thetable{} ... continued from previous page}} \\
\midrule
\textbf{ID} & \textbf{Hamiltonian} & \textbf{Initial State} & \textbf{Target State} \\%& \textbf{System Name} \\
\midrule
\endhead

\multicolumn{4}{r}{{Continued on next page ...}} \\ \midrule
\endfoot

\midrule
\endlastfoot

% ==================== SINGLE QUBIT SYSTEMS ====================
\multicolumn{4}{c}{Single-qubit Systems} \\
\midrule

SQ1 & $H = u(t)(\sigma_x + \sigma_z)$ & $\ket{+}$ & $\ket{0}$ \\%& \seqsplit{SQStaticNoXZ\_IE\_Hdmd} \\
\midrule

SQ2 & $H = \sigma_x + u(t)\sigma_z$ & $\ket{0}$ & $\ket{1}$ \\%& \seqsplit{SQStaticXControlZ\_LZ} \\
\midrule

SQ3 & $H = \sigma_z + u(t)\sigma_x$ & $\ket{0}$ & $\ket{1}$ \\%& \seqsplit{SQStaticZControlX\_LZ} \\
\midrule

SQ4 & $H = \sigma_x + u(t)(\sigma_x + \sigma_z)$ & $\ket{+}$ & $\ket{1}$ \\%& \seqsplit{SQStaticXControlXZ} \\
\midrule

SQ5 & $H = u_y(t)\sigma_y + u_z(t)\sigma_z$ & $\ket{0}$ & $\ket{1}$ \\%& \seqsplit{SQStaticNoControlYZ} \\
\midrule

SQ6 & $H = u_x(t)\sigma_x + u_y(t)\sigma_y + u_z(t)\sigma_z$ & $\ket{0}$ & $\ket{+}$ \\%& \seqsplit{SQStaticNoControlXYZ} \\
\midrule

SQ7 & $H = 0.5\sigma_x + u_x(t)\sigma_x + u_y(t)\sigma_y + u_z(t)\sigma_z$ & $\ket{0}$ & $\ket{1}$ \\%& \seqsplit{SQStaticXControlXYZ} \\
\midrule

SQ8 & $H = 0.5\sigma_y + u_z(t)\sigma_z$ & $\ket{1}$ & $\ket{0}$ \\%& \seqsplit{SQStaticYControlZ} \\
\midrule

SQ9 & $H = 0.5\sigma_y + u_x(t)\sigma_x + u_y(t)\sigma_y + u_z(t)\sigma_z$ & $\ket{1}$ & $\ket{+}$ \\%& \seqsplit{SQStaticYControlXYZ} \\
\midrule

SQ10 & $H = 0.5\sigma_z + u_x(t)\sigma_x$ & $\ket{0}$ & $\ket{1}$ \\%& \seqsplit{SQStaticZControlX} \\
\midrule

SQ11 & $H = 0.5\sigma_z + u_x(t)\sigma_x + u_z(t)\sigma_z$ & $\ket{0}$ & $\ket{1}$ \\%& \seqsplit{SQStaticZControlXZ} \\
\midrule

SQ12 & $H = 0.5\sigma_z + u_x(t)\sigma_x + u_y(t)\sigma_y + u_z(t)\sigma_z$ & $\ket{0}$ & $\ket{+}$ \\%& \seqsplit{SQStaticZControlXYZ} \\
\midrule

SQ13 & $H = 0.5(\sigma_x + \sigma_y) + u_x(t)\sigma_x + u_y(t)\sigma_y$ & $\ket{1}$ & $\ket{-}$ \\%& \seqsplit{SQStaticXYControlXY} \\
\midrule

SQ14 & $H = 0.5(\sigma_x + \sigma_y) + u_x(t)\sigma_x + u_y(t)\sigma_y + u_z(t)\sigma_z$ & $\ket{0}$ & $\frac{\ket{1}-\ket{0}}{\sqrt{2}}$ \\%& \seqsplit{SQStaticXYControlXYZ} \\
\midrule

SQ15 & $H = 0.5(\sigma_x + \sigma_z) + u_z(t)\sigma_z$ & $\ket{0}$ & $\ket{1}$ \\%& \seqsplit{SQStaticXZControlZ} \\
\midrule

SQ16 & $H = 0.5(\sigma_x + \sigma_z) + u_x(t)\sigma_x + u_z(t)\sigma_z$ & $\ket{0}$ & $\ket{+}$ \\%& \seqsplit{SQStaticXZControlXZ} \\
\midrule

SQ17 & $H = 0.5(\sigma_x + \sigma_z) + u_x(t)\sigma_x + u_y(t)\sigma_y + u_z(t)\sigma_z$ & $\ket{1}$ & $\ket{0}$ \\%& \seqsplit{SQStaticXZControlXYZ} \\
\midrule

SQ18 & $H = 0.5(\sigma_y + \sigma_z) + u_x(t)\sigma_x$ & $\ket{0}$ & $\ket{+}$ \\%& \seqsplit{SQStaticYZControlX} \\
\midrule

SQ19 & $H = 0.5(\sigma_y + \sigma_z) + u_x(t)\sigma_x + u_y(t)\sigma_y + u_z(t)\sigma_z$ & $\ket{1}$ & $\ket{0}$ \\%& \seqsplit{SQStaticYZControlXYZ} \\
\midrule

SQ20 & $H = 0.5(\sigma_x + \sigma_y + \sigma_z) + u_x(t)\sigma_x + u_y(t)\sigma_y + u_z(t)\sigma_z$ & $\ket{1}$ & $\ket{+}$ \\%& \seqsplit{SQStaticXYZControlXYZ} \\
\midrule

SQ21 & $H = (\sigma_x + \sigma_y + \sigma_z) + u_x(t)\sigma_x + u_y(t)\sigma_y + u_z(t)\sigma_z$ & $\ket{0}$ & $\ket{1}$ \\%& \seqsplit{SQStaticXYZControlXYZ2} \\
\bottomrule
\end{longtable}

% =========================================================================================
\begin{longtable}{cL{8.250cm}C{1.0cm}C{1.0cm}}
\caption{Details of two-qubit Hamiltonians used in Multi-Task RL Control} \label{tab:two-qubit-hamiltonians} \\

\toprule
\textbf{ID} & \textbf{Hamiltonian } & \textbf{Initial State} & \textbf{Target State} \\%& \textbf{System Name} \\
\toprule
\endfirsthead

\multicolumn{4}{c}%
{{\tablename\ \thetable{} ... continued from previous page}} \\
\midrule
\textbf{ID} & \textbf{Hamiltonian} & \textbf{Initial State} & \textbf{Target State} \\%& \textbf{System Name} \\
\midrule
\endhead

\multicolumn{4}{r}{{Continued on next page ...}} \\ \midrule
\endfoot

\midrule
\endlastfoot
% ==================== TWO-QUBIT XX SYSTEMS ====================
\multicolumn{4}{c}{Two-qubit Systems} \\
\midrule

TQ1 & $H = \sigma_x^1\sigma_x^2 + u_x(t)\sigma_x^1 + v_z(t)\sigma_z^2$ & $\ket{00}$ & $\ket{\Phi^+}$ \\%& \seqsplit{TQInterXXControlXZ} \\
\midrule

TQ2 & $H = \sigma_x^1\sigma_x^2 + u_x(t)\sigma_x^1 + v_x(t)\sigma_x^2 + v_y(t)\sigma_y^2$ & $\ket{00}$ & $\ket{\Phi^+}$ \\%& \seqsplit{TQInterXXControlXY} \\
\midrule

TQ3 & $H = \sigma_x^1 \sigma_x^2 + u_y(t)\sigma_y^1 + \sum_{\alpha \in \{x,y,z\}} v_\alpha(t)\sigma_\alpha^2$ & $\ket{00}$ & $\ket{\Psi^+}$ \\%& \seqsplit{TQInterXXControlYXYZ} \\
\midrule

TQ4 & $H = \sigma_x^1 \sigma_x^2 + \sum_{\alpha \in \{x,y\}} u_\alpha(t)\sigma_\alpha^1 + \sum_{\alpha \in \{x,y,z\}} v_\alpha(t)\sigma_\alpha^2$ & $\ket{00}$ & $\ket{\Psi^+}$ \\%& \seqsplit{TQInterXXControlXYXYZ} \\
\midrule

TQ5 & $H = \sigma_x^1 \sigma_x^2 + \sum_{\alpha \in \{y,z\}} u_\alpha(t)\sigma_\alpha^1 + \sum_{\alpha \in \{x,y,z\}} v_\alpha(t)\sigma_\alpha^2$ & $\ket{01}$ & $\ket{11}$ \\%& \seqsplit{TQInterXXControlYZXYZ} \\
\midrule

TQ6 & $H = \sigma_x^1 \sigma_x^2 + \sum_{\alpha \in \{x,y,z\}} \left( u_\alpha(t)\sigma_\alpha^1 + v_\alpha(t)\sigma_\alpha^2 \right)$ & $\ket{00}$ & $\ket{\Phi^+}$ \\%& \seqsplit{TQInterXXControlXYZXYZ} \\
\midrule
% ==================== TWO-QUBIT YY SYSTEMS ====================

TQ7 & $H = \sigma_y^1 \sigma_y^2 + \sum_{\alpha \in \{x,z\}} u_\alpha(t)\sigma_\alpha^1 + \sum_{\alpha \in \{x,y,z\}} v_\alpha(t)\sigma_\alpha^2$ & $\ket{00}$ & $\ket{\Phi^+}$ \\%& \seqsplit{TQInterYYControlXZXYZ} \\
\midrule

TQ8 & $H = \sigma_y^1 \sigma_y^2 + \sum_{\alpha \in \{y,z\}} u_\alpha(t)\sigma_\alpha^1 + \sum_{\alpha \in \{x,y,z\}} v_\alpha(t)\sigma_\alpha^2$ & $\ket{01}$ & $\frac{\ket{10}+\ket{11}}{\sqrt{2}}$ \\%& \seqsplit{TQInterYYControlYZXYZ} \\
\midrule

TQ9 & $H = \sigma_y^1 \sigma_y^2 + \sum_{\alpha \in \{x,y,z\}} \left( u_\alpha(t)\sigma_\alpha^1 + v_\alpha(t)\sigma_\alpha^2 \right)$ & $\ket{\Psi^+}$ & $\ket{00}$ \\%& \seqsplit{TQInterYYControlXYZXYZ} \\
\midrule
% ==================== TWO-QUBIT ZZ SYSTEMS ====================
TQ10 & $H = \sigma_z^1 \sigma_z^2 + u_y(t)\sigma_y^1 + \sum_{\alpha \in \{x,y,z\}} v_\alpha(t)\sigma_\alpha^2$ & $\ket{\Psi^+}$ & $\ket{\Phi^+}$ \\%& \seqsplit{TQInterZZControlYXYZ} \\
\midrule

TQ11 & $H = \sigma_z^1 \sigma_z^2 + \sum_{\alpha \in \{x,y\}} u_\alpha(t)\sigma_\alpha^1 + \sum_{\alpha \in \{x,y,z\}} v_\alpha(t)\sigma_\alpha^2$ & $\ket{00}$ & $\ket{\Phi^+}$ \\%& \seqsplit{TQInterZZControlXYXYZ} \\
\midrule

TQ12 & $H = \sigma_z^1 \sigma_z^2 + \sum_{\alpha \in \{x,y,z\}} \left( u_\alpha(t)\sigma_\alpha^1 + v_\alpha(t)\sigma_\alpha^2 \right)$ & $\ket{00}$ & $\ket{\Phi^-}$ \\%& \seqsplit{TQInterZZControlXYZXYZ} \\
\midrule
% ==================== TWO-QUBIT XY MODEL ====================
TQ13 & $H = \sum_{\alpha \in \{x,y\}} \sigma_\alpha^1 \sigma_\alpha^2 + u_x(t)\sigma_x^1 + v_y(t)\sigma_y^2$ & $\ket{\Phi^+}$ & $\ket{\Psi^-}$ \\%& \seqsplit{TQInterXYControlXY} \\
\midrule

TQ14 & $H = \sum_{\alpha \in \{x,y\}} \sigma_\alpha^1 \sigma_\alpha^2 + \sum_{\alpha \in \{x,y\}} u_\alpha(t)\sigma_\alpha^1 + \sum_{\alpha \in \{x,y,z\}} v_\alpha(t)\sigma_\alpha^2$ & $\ket{00}$ & $\ket{\Psi^+}$ \\
\midrule

TQ15 & $H = \sum_{\alpha \in \{x,y\}} \sigma_\alpha^1 \sigma_\alpha^2 + \sum_{\alpha \in \{x,y,z\}} \left( u_\alpha(t)\sigma_\alpha^1 + v_\alpha(t)\sigma_\alpha^2 \right)$ & $\ket{11}$ & $\ket{\Phi^-}$ \\
\midrule
% ==================== TWO-QUBIT XZ MODEL ====================
TQ16 & $H = \sum_{\alpha \in \{x,z\}} \sigma_\alpha^1 \sigma_\alpha^2 + \sum_{\alpha \in \{x,z\}} u_\alpha(t)\sigma_\alpha^1 + \sum_{\alpha \in \{x,y,z\}} v_\alpha(t)\sigma_\alpha^2$ & $\ket{00}$ & $\ket{\Psi^+}$ \\
\midrule

TQ17 & $H = \sum_{\alpha \in \{x,z\}} \sigma_\alpha^1 \sigma_\alpha^2 + \sum_{\alpha \in \{x,y,z\}} u_\alpha(t)\sigma_\alpha^1 + \sum_{\alpha \in \{x,y,z\}} v_\alpha(t)\sigma_\alpha^2$ & $\ket{\Phi^-}$ & $\ket{\Psi^+}$ \\
\midrule
% ==================== TWO-QUBIT YZ MODEL ====================
TQ18 & $H = \sum_{\alpha \in \{y,z\}} \sigma_\alpha^1 \sigma_\alpha^2 + \sum_{\alpha \in \{x,z\}} u_\alpha(t)\sigma_\alpha^1 + \sum_{\alpha \in \{x,y,z\}} v_\alpha(t)\sigma_\alpha^2$ & $\ket{00}$ & $\ket{\Psi^+}$ \\
\midrule

TQ19 & $H = \sum_{\alpha \in \{y,z\}} \sigma_\alpha^1 \sigma_\alpha^2 + \sum_{\alpha \in \{x,y,z\}} u_\alpha(t)\sigma_\alpha^1 + \sum_{\alpha \in \{x,y,z\}} v_\alpha(t)\sigma_\alpha^2$ & $\ket{00}$ & $\ket{\Phi^+}$ \\
\midrule
% ==================== TWO-QUBIT HEISENBERG ====================
TQ20 & $H = \sum_{\alpha \in \{x,y,z\}} \sigma_\alpha^1 \sigma_\alpha^2 + \sum_{\alpha \in \{x,y,z\}} v_\alpha(t)\sigma_\alpha^2$ & $\ket{\Phi^+}$ & $\ket{\Psi^-}$ \\
\midrule

TQ21 & $H = \sum_{\alpha \in \{x,y,z\}} \sigma_\alpha^1 \sigma_\alpha^2 + u_x(t)\sigma_x^1 + \sum_{\alpha \in \{y,z\}} v_\alpha(t)\sigma_\alpha^2$ & $\ket{00}$ & $\ket{\Psi^-}$ \\
\midrule

TQ22 & $H = \sum_{\alpha \in \{x,y,z\}} \sigma_\alpha^1 \sigma_\alpha^2 + \sum_{\alpha \in \{y,z\}} u_\alpha(t)\sigma_\alpha^1 + \sum_{\alpha \in \{x,y,z\}} v_\alpha(t)\sigma_\alpha^2$ & $\ket{\Psi^-}$ & $\ket{\Phi^+}$ \\%& \seqsplit{TQHeisenbergXYZControlYZXYZ} \\
\midrule

TQ23 & $H = \sum_{\alpha \in \{x,y,z\}} \sigma_\alpha^1 \sigma_\alpha^2 + \sum_{\alpha \in \{x,y,z\}} u_\alpha(t)\sigma_\alpha^1 + \sum_{\alpha \in \{x,z\}} v_\alpha(t)\sigma_\alpha^2$ & $\ket{00}$ & $\ket{\Phi^+}$ \\
\midrule

TQ24 & $H = \sum_{\alpha \in \{x,y,z\}} \sigma_\alpha^1 \sigma_\alpha^2 + \sum_{\alpha \in \{x,y,z\}} u_\alpha(t)\sigma_\alpha^1 + \sum_{\alpha \in \{x,y,z\}} v_\alpha(t)\sigma_\alpha^2$ & $\ket{11}$ & $\ket{\Phi^+}$ \\
\midrule
% ==================== TWO-QUBIT SPECIAL ====================
TQ25 & $H = -\sigma_z^1\sigma_z^2 - \frac{1}{2}\sum_{i=1}^{2} \sigma_z^i - \frac{u(t)}{2}\sum_{i=1}^{2} \sigma_x^i$ & $\ket{++}$ & $\ket{00}$ \\
\midrule

TQ26 & $H = \frac{1}{2} \sum_{\alpha \in \{x,y\}} \sigma_\alpha^1 \sigma_\alpha^2 + \frac{1}{2} \sum_{i=1}^{2} \left( \sigma_z^i + u_i(t)\sigma_x^i \right)$ & $\ket{00}$ & $\ket{\Phi^+}$ \\

\bottomrule
\end{longtable}

% =======================================================================
\begin{longtable}{cL{8.250cm}C{1.0cm}C{1.0cm}}
\caption{Details of three-qubit Hamiltonians used in Multi-Task RL Control} \label{tab:three-qubit-hamiltonians} \\

\toprule
\textbf{ID} & \textbf{Hamiltonian } & \textbf{Initial State} & \textbf{Target State} \\%& \textbf{System Name} \\
\toprule
\endfirsthead

\multicolumn{4}{c}%
{{\tablename\ \thetable{} ... continued from previous page}} \\
\midrule
\textbf{ID} & \textbf{Hamiltonian} & \textbf{Initial State} & \textbf{Target State} \\%& \textbf{System Name} \\
\midrule
\endhead

\multicolumn{4}{r}{{Continued on next page ...}} \\ \midrule
\endfoot

\midrule
\endlastfoot
% ==================== THREE-QUBIT ====================
\multicolumn{4}{c}{Three-qubit Systems} \\
\midrule

ThQ1 & $H = u_1(t)\sigma_z^1\sigma_x^2 + u_2(t)\sigma_x^2\sigma_z^3 + u_3(t)\sigma_x^2 + u_4(t)\sigma_y^2$ & $\ket{000}$ & $-i\ket{010}$ \\%& \seqsplit{ThQiTOffoli} \\
\midrule

ThQ2 & $H=(\sigma_z^1\sigma_z^2+\sigma_z^1\sigma_z^3+\sigma_z^2\sigma_z^3) +  \sum_{\alpha \in \{x,y,z\}} u_\alpha(t) \sum_{k=1}^{3} \sigma_\alpha^k$ & $\ket{000}$ & $\ket{D_3^1}$ \\
\midrule

ThQ3 & $H = u_{12}(t)(\sigma_x^1\sigma_x^2+\sigma_y^1\sigma_y^2) + u_{23}(t)(\sigma_x^2\sigma_x^3+\sigma_y^2\sigma_y^3)$ & $\ket{100}$ & $\ket{001}$ \\%& \seqsplit{ThQ\_XY\_exchange} \\
\midrule

ThQ4 & $H = u_1(t)(\sigma_z^1\sigma_z^2+\sigma_z^2\sigma_z^3+\sigma_z^1\sigma_z^3) + u_2(t)\sum_{k=1}^3\sigma_x^k + u_3(t)\sum_{k=1}^3\sigma_z^k$ & $\ket{+++}$ & $\ket{\text{GHZ}}$ \\%& \seqsplit{ThQGHZCollectiveControl} \\
\bottomrule
\end{longtable}

\vspace{0.5em}
\noindent\textbf{Notation:}
\begin{itemize}
    \item $\ket{+} = (\ket{0}+\ket{1})/\sqrt{2}$
    \item $\ket{-} = (\ket{0}-\ket{1})/\sqrt{2}$
    \item $\ket{\Phi^+} = (\ket{00}+\ket{11})/\sqrt{2}$
    \item $\ket{\Phi^-} = (\ket{00}-\ket{11})/\sqrt{2}$
    \item $\ket{\Psi^+} = (\ket{01}+\ket{10})/\sqrt{2}$
    \item $\ket{\Psi^-} = (\ket{01}-\ket{10})/\sqrt{2}$
    \item $\ket{++} = (\ket{00}+\ket{01}+\ket{10}+\ket{11})/\sqrt{4}$
    \item $\ket{\text{D}_3^1} = (\ket{100}+\ket{010}+\ket{001})/\sqrt{3}$
    \item $\ket{\text{GHZ}} = (\ket{000}+\ket{111})/\sqrt{2}$ 
    \item $\ket{+++} = (\ket{000} + \ket{001} + \ket{010} + \ket{011} + \ket{100} + \ket{101} + \ket{110} + \ket{111})/\sqrt{8}$
    \item Superscripts denote qubit indices. $u_\alpha(t), v_\alpha(t)$ are time-dependent control fields.
\end{itemize}

\FloatBarrier

\section{Sample Results of the Hamiltonian Expansion Experiment} \label{appendix:B}
Results of experiments on held-out samples for models trained with 10 Hamiltonians and 20 Hamiltonians is presented below. The results for other group of Hamiltonians are not different in a meaningful way, so we present only the two plots for reference. Training on 10 Hamiltonians yields fidelities $>0.95$ for most single-qubit (SQ) systems, while two-qubit (TQ) and three-qubit (ThQ) systems reach up to $0.90$ (averaging $0.55$–$0.75$).

\begin{figure}[htbp]
     \centering
     \begin{subfigure}[b]{0.99\textwidth}
         \centering
         \includegraphics[width=0.999\textwidth]{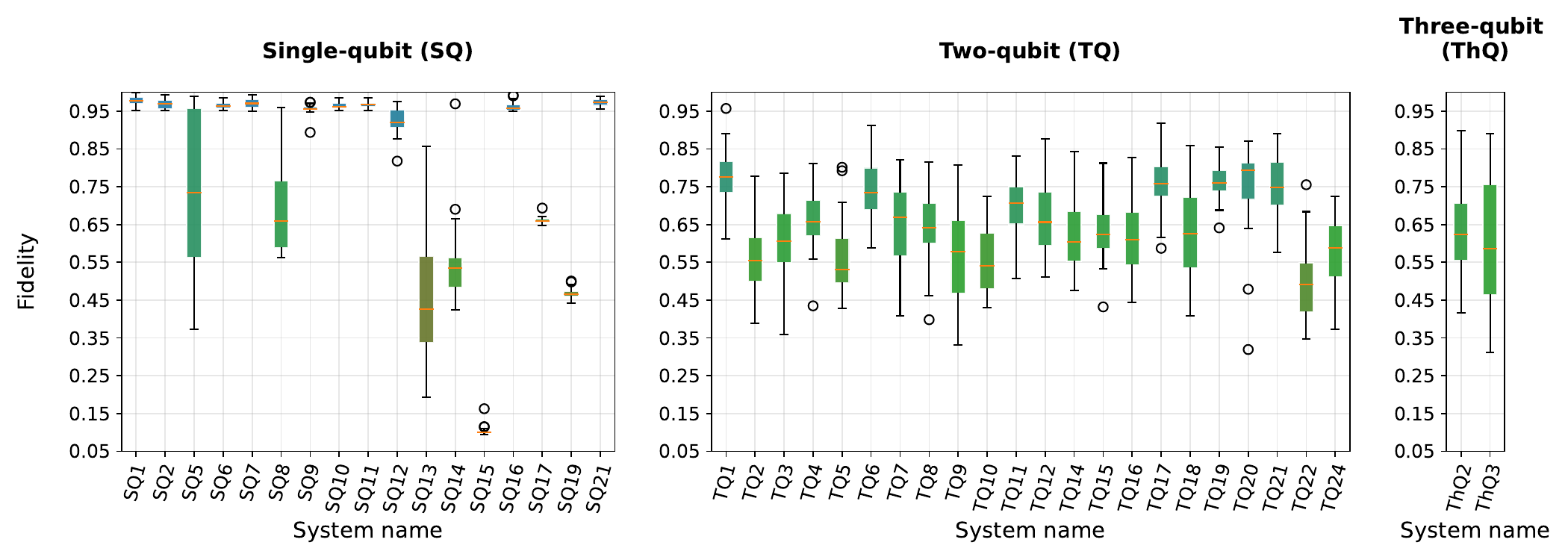}
         \caption{Fidelity distribution on held-out Hamiltonians for model trained with 10 Hamiltonians}
         \label{fig:sub_10_hamiltonians}
     \end{subfigure}
     \hfill
     \begin{subfigure}[b]{0.99\textwidth}
         \centering
        \includegraphics[width=0.999\textwidth]{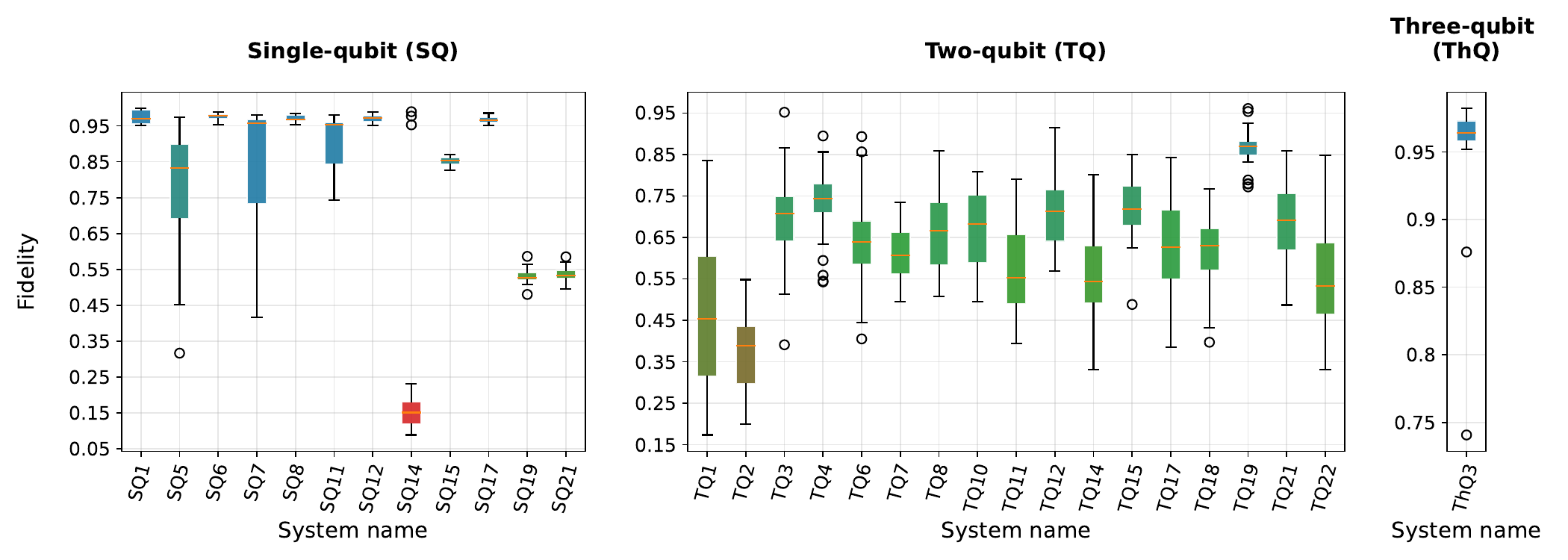}
         \caption{Fidelity distribution on held-out Hamiltonians for model trained with 20 Hamiltonians}
         \label{fig:sub_20_hamiltonians}
     \end{subfigure}
     
     \caption{Hamiltonian expansion results}
     \label{fig:main_comparison}
\end{figure}
Expanding to 20 training systems maintains high SQ performance, stabilizes TQ fidelities between $0.65$–$0.75$, and boosts the ThQ fidelity to approximately $0.95$. Notably, these results are obtained from held-out Hamiltonians, demonstrating the agent's ability to generalize to entirely unseen systems. 
\end{appendices}

%%===========================================================================================%%
%% If you are submitting to one of the Nature Portfolio journals, using the eJP submission   %%
%% system, please include the references within the manuscript file itself. You may do this  %%
%% by copying the reference list from your .bbl file, paste it into the main manuscript .tex %%
%% file, and delete the associated \verb+\bibliography+ commands.                            %%
%%===========================================================================================%%
\FloatBarrier
\bibliography{references}% 

@article{preskill2018quantum,
  title={Quantum computing in the NISQ era and beyond},
  author={Preskill, John},
  journal={Quantum},
  volume={2},
  pages={79},
  year={2018},
  url={https://doi.org/10.22331/q-2018-08-06-79},
  doi={https://doi.org/10.22331/q-2018-08-06-79},
  publisher={Verein zur F{\"o}rderung des Open Access Publizierens in den Quantenwissenschaften}
}

@article{khaneja2005optimal,
    title = {Optimal control of coupled spin dynamics: design of NMR pulse sequences by gradient ascent algorithms},
    journal = {Journal of Magnetic Resonance},
    volume = {172},
    number = {2},
    pages = {296-305},
    year = {2005},
    issn = {1090-7807},
    doi = {https://doi.org/10.1016/j.jmr.2004.11.004},
    url = {https://doi.org/10.1016/j.jmr.2004.11.004},
    author = {Navin Khaneja and Timo Reiss and Cindie Kehlet and Thomas Schulte-Herbrüggen and Steffen J. Glaser}
}

@article{bukov2018reinforcement,
   title={Reinforcement Learning in Different Phases of Quantum Control},
   volume={8},
   ISSN={2160-3308},
   url={https://dx.doi.org/10.1103/PhysRevX.8.031086},
   DOI={https://dx.doi.org/10.1103/physrevx.8.031086},
   number={3},
   journal={Physical Review X},
   publisher={American Physical Society (APS)},
   author={Bukov, Marin and Day, Alexandre G.bR. and Sels, Dries and Weinberg, Phillip and Polkovnikov, Anatoli and Mehta, Pankaj},
   year={2018},
   month=sep }

@article{zhang2019reinforcement,
  title={When does reinforcement learning stand out in quantum control? {A} comparative study on state preparation},
  author={Zhang, Xiao-Ming and Wei, Zhenyu and Asad, Raza and Yang, Xu-Chen and Wang, Xin},
  journal={npj Quantum Information},
  volume={5},
  number={1},
  pages={85},
  year={2019},
  doi={https://doi.org/10.1038/s41534-019-0201-8},
  url={https://doi.org/10.1038/s41534-019-0201-8},
  publisher={Nature Publishing Group}
}

@article{niu2019universal,
  title={Universal quantum control through deep reinforcement learning},
  author={Niu, Murphy Yuezhen and Boixo, Sergio and Smelyanskiy, Vadim and Neven, Hartmut},
  journal={npj Quantum Information},
  volume={5},
  number={1},
  pages={33},
  year={2019},
  doi={https://doi.org/10.1038/s41534-019-0141-3},
  url={https://doi.org/10.1038/s41534-019-0141-3},
  publisher={Nature Publishing Group}
}

@article{lindblad1976generators,
  title={On the generators of quantum dynamical semigroups},
  author={Lindblad, G{\"o}ran},
  journal={Communications in Mathematical Physics},
  volume={48},
  number={2},
  pages={119--130},
  year={1976},
  doi={https://doi.org/10.1007/BF01608499},
  url={https://doi.org/10.1007/BF01608499},
  publisher={Springer}
}

@article{zhang2025meta,
	author = {Zhang, Shihui and Miao, Zibo and Pan, Yu and Tao, Sibo and Chen, Yu},
	date = {2025/05/19},
	doi = {10.1038/s41534-025-01034-9},
	id = {Zhang2025},
	isbn = {2056-6387},
	journal = {npj Quantum Information},
	number = {1},
	pages = {81},
	title = {Meta-learning assisted robust control of universal quantum gates with uncertainties},
	url = {https://doi.org/10.1038/s41534-025-01034-9},
	volume = {11},
	year = {2025}
}

@article{PhysRevLett.132.193801,
  title = {Universally Robust Quantum Control},
  author = {Poggi, Pablo M. and De Chiara, Gabriele and Campbell, Steve and Kiely, Anthony},
  journal = {Phys. Rev. Lett.},
  volume = {132},
  issue = {19},
  pages = {193801},
  numpages = {6},
  year = {2024},
  month = {May},
  publisher = {American Physical Society},
  doi = {10.1103/PhysRevLett.132.193801},
  url = {https://link.aps.org/doi/10.1103/PhysRevLett.132.193801}
}

@book{Nielsen_Chuang_2010, 
    place={Cambridge}, 
    title={Quantum Computation and Quantum Information: 10th Anniversary Edition}, 
    publisher={Cambridge University Press}, 
    author={Nielsen, Michael A. and Chuang, Isaac L.}, 
    url={https://doi.org/10.1017/CBO9780511976667},
    doi={https://doi.org/10.1017/CBO9780511976667},
    year={2010}
}

@article{manzano_2020,
  author  = {Manzano, Daniel},
  title   = {A short introduction to the {Lindblad} master equation},
  journal = {AIP Advances},
  volume  = {10},
  number  = {2},
  pages   = {025106},
  year    = {2020},
  url     = {https://doi.org/10.1063/1.5115323},
  doi     = {https://doi.org/10.1063/1.5115323}
}

@article{Ball_2021,
  author  = {Ball, Harrison and Biercuk, Michael J. and Carvalho, Andr{\'e} R. R. and Chen, Jiayin and Hush, Michael and De Castro, Leonardo A. and Li, Li and Liebermann, Per J. and Slatyer, Harry J. and Edmunds, Claire and others},
  title   = {Software tools for quantum control: improving quantum computer performance through noise and error suppression},
  journal = {Quantum Science and Technology},
  volume  = {6},
  number  = {4},
  pages   = {044011},
  doi     = {10.1088/2058-9565/abdca6},
  url = {https://doi.org/10.1088/2058-9565/abdca6},
  year = {2021},
  month = {sep},
  publisher = {IOP Publishing},
}

@article{Ian_Rabitz_2003_quantum_control,
    title = {Quantum physics under control},
    author = {Ian Walmsley and Herschel Rabitz},
    year = {2003},
    month = {aug},
    doi = {https://doi.org/10.1063/1.1611352},
    url = {https://doi.org/10.1063/1.1611352},
    language = {English (US)},
    volume = {56},
    pages = {43--49},
    journal = {Physics Today},
    issn = {0031-9228},
    publisher = {American Institute of Physics},
}

@book{BreuerPetruccione2002,
  title     = {The Theory of Open Quantum Systems},
  author    = {Breuer, Heinz-Peter and Petruccione, Francesco},
  year      = {2002},
  publisher = {Oxford University Press},
  address   = {Oxford, UK},
  isbn = {9780199213900},
  doi = {10.1093/acprof:oso/9780199213900.001.0001},
  url = {https://doi.org/10.1093/acprof:oso/9780199213900.001.0001},
}

@article{barnes2012analytically,
  title={Analytically solvable driven time-dependent two-level quantum systems},
  author={Barnes, Edwin and Das Sarma, Sankar},
  journal={Physical review letters},
  volume={109},
  number={6},
  pages={060401},
  year={2012},
  month = {Aug},
  publisher = {American Physical Society},
  doi = {https://link.aps.org/doi/10.1103/PhysRevLett.109.060401},
  url = {https://link.aps.org/doi/10.1103/PhysRevLett.109.060401}
}

@article{kosut2013robust,
  title = {Robust control of quantum gates via sequential convex programming},
  author = {Kosut, Robert L. and Grace, Matthew D. and Brif, Constantin},
  journal = {Phys. Rev. A},
  volume = {88},
  issue = {5},
  pages = {052326},
  numpages = {12},
  year = {2013},
  month = {Nov},
  publisher = {American Physical Society},
  doi = {https://link.aps.org/doi/10.1103/PhysRevA.88.052326},
  url = {https://link.aps.org/doi/10.1103/PhysRevA.88.052326}
}

@article{lin2020time,
  title={Time-optimal control of a dissipative qubit},
  author={Lin, Chungwei and Sels, Dries and Wang, Yebin},
  journal={Physical Review A},
  volume={101},
  number={2},
  pages={022320},
  year={2020},
  publisher={APS},
  doi = {https://link.aps.org/doi/10.1103/PhysRevA.101.022320},
  url = {https://link.aps.org/doi/10.1103/PhysRevA.101.022320}
}

@article{del2012assisted,
   title={Assisted Finite-Rate Adiabatic Passage Across a Quantum Critical Point: Exact Solution for the Quantum Ising Model},
   volume={109},
   ISSN={1079-7114},
   url={http://dx.doi.org/10.1103/PhysRevLett.109.115703},
   DOI={http://dx.doi.org/10.1103/physrevlett.109.115703},
   number={11},
   journal={Physical Review Letters},
   publisher={American Physical Society (APS)},
   author={del Campo, Adolfo and Rams, Marek M. and Zurek, Wojciech H.},
   year={2012},
   month=sep }

@article{Xikuns4159802341688z,
	author = {Li, Xikun},
	date = {2023/09/07},
	doi = {10.1038/s41598-023-41688-z},
	id = {Li2023},
	isbn = {2045-2322},
	journal = {Scientific Reports},
	number = {1},
	pages = {14734},
	title = {Optimal control of quantum state preparation and entanglement creation in two-qubit quantum system with bounded amplitude},
	url = {https://doi.org/10.1038/s41598-023-41688-z},
	volume = {13},
	year = {2023}
}

@article{Chen2025,
	author = {Chen, Zhen and Liu, Weiyang and Ma, Yanjun and Sun, Weijie and Wang, Ruixia and Wang, He and Xu, Huikai and Xue, Guangming and Yan, Haisheng and Yang, Zhen and Ding, Jiayu and Gao, Yang and Li, Feiyu and Zhang, Yujia and Zhang, Zikang and Jin, Yirong and Yu, Haifeng and Chen, Jianxin and Yan, Fei},
	date = {2025/09/01},
	doi = {https://doi.org/10.1038/s41567-025-02990-x},
	id = {Chen2025},
	isbn = {1745-2481},
	journal = {Nature Physics},
	number = {9},
	pages = {1489--1496},
	title = {Efficient implementation of arbitrary two-qubit gates using unified control},
	url = {https://doi.org/10.1038/s41567-025-02990-x},
	volume = {21},
	year = {2025}
}

@article{GoerzChristiane2015,
   title={Hybrid optimization schemes for quantum control},
   volume={2},
   ISSN={2196-0763},
   url={http://dx.doi.org/10.1140/epjqt/s40507-015-0034-0},
   DOI={http://dx.doi.org/10.1140/epjqt/s40507-015-0034-0},
   number={1},
   journal={EPJ Quantum Technology},
   publisher={Springer Science and Business Media LLC},
   author={Goerz, Michael H and Whaley, K Birgitta and Koch, Christiane P},
   year={2015},
   month=sep 
}

@InProceedings{haarnoja2018soft,
  title = 	 {Soft Actor-Critic: Off-Policy Maximum Entropy Deep Reinforcement Learning with a Stochastic Actor},
  author =       {Haarnoja, Tuomas and Zhou, Aurick and Abbeel, Pieter and Levine, Sergey},
  booktitle = 	 {Proceedings of the 35th International Conference on Machine Learning},
  pages = 	 {1861--1870},
  year = 	 {2018},
  editor = 	 {Dy, Jennifer and Krause, Andreas},
  volume = 	 {80},
  series = 	 {Proceedings of Machine Learning Research},
  month = 	 {10--15 Jul},
  publisher =    {PMLR},
  url = 	 {https://proceedings.mlr.press/v80/haarnoja18b.html},
}

@article{Haarnoja2018SoftAA,
  title={Soft Actor-Critic Algorithms and Applications},
  author={Tuomas Haarnoja and Aurick Zhou and Kristian Hartikainen and G. Tucker and Sehoon Ha and Jie Tan and Vikash Kumar and Henry Zhu and Abhishek Gupta and P. Abbeel and Sergey Levine},
  journal={ArXiv},
  year={2018},
  volume={abs/1812.05905},
  url={https://api.semanticscholar.org/CorpusID:55703664}
}

@misc{oneil2024robustnessdynamicquantumcontrol,
      title={Robustness of Dynamic Quantum Control: Differential Sensitivity Bound}, 
      author={S. P. O'Neil and C. A. Weidner and E. A. Jonckheere and F. C. Langbein and S. G. Schirmer},
      year={2024},
      eprint={2401.00301},
      archivePrefix={arXiv},
      primaryClass={quant-ph},
      url={https://arxiv.org/abs/2401.00301}, 
}

@article{PhysRevA.107.032606,
  title = {Statistically characterizing robustness and fidelity of quantum controls and quantum control algorithms},
  author = {Khalid, Irtaza and Weidner, Carrie A. and Jonckheere, Edmond A. and Shermer, Sophie G. and Langbein, Frank C.},
  journal = {Phys. Rev. A},
  volume = {107},
  issue = {3},
  pages = {032606},
  numpages = {22},
  year = {2023},
  month = {Mar},
  publisher = {American Physical Society},
  doi = {https://link.aps.org/doi/10.1103/PhysRevA.107.032606},
  url = {https://link.aps.org/doi/10.1103/PhysRevA.107.032606}
}

@article{dong2015sampling,
   author = {Dong, Daoyi and Mabrok, Mohamed A. and Petersen, Ian R. and Qi, Bo and Chen, Chunlin and Rabitz, Herschel},
   title = {Sampling-based learning control for quantum systems with uncertainties},
   journal = {IEEE Transactions on Control Systems Technology},
   volume = {23},
   number = {6},
   pages = {2155--2166},
   year = {2015},
   doi = {https://doi.org/10.1109/TCST.2015.2404292},
   url = {https://doi.org/10.1109/TCST.2015.2404292}
 }

@article{Koswara_2014,
   title={Robustness of controlled quantum dynamics},
   volume={90},
   ISSN={1094-1622},
   url={http://dx.doi.org/10.1103/PhysRevA.90.043414},
   DOI={http://dx.doi.org/10.1103/physreva.90.043414},
   number={4},
   journal={Physical Review A},
   publisher={American Physical Society (APS)},
   author={Koswara, Andy and Chakrabarti, Raj},
   year={2014},
   month=oct }

@article{PhysRevA.85.052313,
  title = {Optimized pulses for the control of uncertain qubits},
  author = {Grace, Matthew D. and Dominy, Jason M. and Witzel, Wayne M. and Carroll, Malcolm S.},
  journal = {Phys. Rev. A},
  volume = {85},
  issue = {5},
  pages = {052313},
  numpages = {15},
  year = {2012},
  month = {May},
  publisher = {American Physical Society},
  doi = {https://link.aps.org/doi/10.1103/PhysRevA.85.052313},
  url = {https://link.aps.org/doi/10.1103/PhysRevA.85.052313}
}

@article{stable-baselines3,
  author  = {Antonin Raffin and Ashley Hill and Adam Gleave and Anssi Kanervisto and Maximilian Ernestus and Noah Dormann},
  title   = {Stable-Baselines3: Reliable Reinforcement Learning Implementations},
  journal = {Journal of Machine Learning Research},
  year    = {2021},
  volume  = {22},
  number  = {268},
  pages   = {1-8},
  url     = {http://jmlr.org/papers/v22/20-1364.html}
}

@misc{nair2010rectified,
  author       = {Nair, Vinod and Hinton, Geoffrey E.},
  title        = {Rectified Linear Units Improve Restricted Boltzmann Machines},
  howpublished = {In \textit{Proceedings of the 27th International Conference on Machine Learning (ICML)}},
  year         = {2010},
  pages        = {807-814},
  note         = {Haifa, Israel}
}

@article{Kim2022,
	author = {Kim, Yosep and Morvan, Alexis and Nguyen, Long B. and Naik, Ravi K. and J{\"u}nger, Christian and Chen, Larry and Kreikebaum, John Mark and Santiago, David I. and Siddiqi, Irfan},
	date = {2022/07/01},
	doi = {https://doi.org/10.1038/s41567-022-01590-3},
	id = {Kim2022},
	isbn = {1745-2481},
	journal = {Nature Physics},
	number = {7},
	pages = {783--788},
	title = {High-fidelity three-qubit iToffoli gate for fixed-frequency superconducting qubits},
	url = {https://doi.org/10.1038/s41567-022-01590-3},
	volume = {18},
	year = {2022}
}

@article{stojanovic2023dicke,
  title={Dicke-state preparation through global transverse control of Ising-coupled qubits},
  author={Stojanovi{\'c}, Vladimir M and Nauth, Julian K},
  journal={Physical Review A},
  url={http://dx.doi.org/10.1103/PhysRevA.108.012608},
  DOI={http://dx.doi.org/10.1103/physreva.108.012608},
  volume={108},
  number={1},
  pages={012608},
  year={2023},
  publisher={APS}
}

@article{Xu2016,
	author = {Xu, Peng and Yang, Xu-Chen and Mei, Feng and Xue, Zheng-Yuan},
	date = {2016/01/25},
	doi = {https://doi.org/10.1038/srep18695},
	id = {Xu2016},
	isbn = {2045-2322},
	journal = {Scientific Reports},
	number = {1},
	pages = {18695},
	title = {Controllable high-fidelity quantum state transfer and entanglement generation in circuit QED},
	url = {https://doi.org/10.1038/srep18695},
	volume = {6},
	year = {2016}
}

@article{brockman2016openai,
  title={Openai gym},
  author={Brockman, Greg and Cheung, Vicki and Pettersson, Ludwig and Schneider, Jonas and Schulman, John and Tang, Jie and Zaremba, Wojciech},
  journal={arXiv preprint arXiv:1606.01540},
  url = {https://doi.org/10.48550/arXiv.1606.01540},
  doi = {https://doi.org/10.48550/arXiv.1606.01540},
  year={2016}
}

@misc{weidner2024robustquantumcontrolclosed,
      title={Robust Quantum Control in Closed and Open Systems: Theory and Practice}, 
      author={C. A. Weidner and E. A. Reed and J. Monroe and B. Sheller and S. O'Neil and E. Maas and E. A. Jonckheere and F. C. Langbein and S. G. Schirmer},
      year={2024},
      eprint={2401.00294},
      archivePrefix={arXiv},
      primaryClass={quant-ph},
      url={https://arxiv.org/abs/2401.00294}, 
}

@misc{finn2017model,
      title={Model-Agnostic Meta-Learning for Fast Adaptation of Deep Networks}, 
      author={Chelsea Finn and Pieter Abbeel and Sergey Levine},
      year={2017},
      eprint={1703.03400},
      archivePrefix={arXiv},
      primaryClass={cs.LG},
      url={https://arxiv.org/abs/1703.03400}, 
}

@article{leclerc2026does,
  title={When Does Adaptation Win? Scaling Laws for Meta-Learning in Quantum Control},
  author={Leclerc, Nima and Miller, Chris and Brawand, Nicholas},
  journal={arXiv preprint arXiv:2601.18973},
  url = {https://arxiv.org/pdf/2601.18973},
  year={2026}
}

@misc{dasgupta2023adaptivemitigationtimevaryingquantum,
      title={Adaptive mitigation of time-varying quantum noise}, 
      author={Samudra Dasgupta and Arshag Danageozian and Travis S. Humble},
      year={2023},
      eprint={2308.14756},
      archivePrefix={arXiv},
      primaryClass={quant-ph},
      url={https://arxiv.org/abs/2308.14756}, 
}

@article{PhysRevLett_121_090502,
  title = {Fluctuations of Energy-Relaxation Times in Superconducting Qubits},
  author = {Klimov, P. V. and Kelly, J. and Chen, Z. and Neeley, M. and Megrant, A. and Burkett, B. and Barends, R. and Arya, K. and Chiaro, B. and Chen, Yu and Dunsworth, A. and Fowler, A. and Foxen, B. and Gidney, C. and Giustina, M. and Graff, R. and Huang, T. and Jeffrey, E. and Lucero, Erik and Mutus, J. Y. and Naaman, O. and Neill, C. and Quintana, C. and Roushan, P. and Sank, Daniel and Vainsencher, A. and Wenner, J. and White, T. C. and Boixo, S. and Babbush, R. and Smelyanskiy, V. N. and Neven, H. and Martinis, John M.},
  journal = {Phys. Rev. Lett.},
  volume = {121},
  issue = {9},
  pages = {090502},
  numpages = {5},
  year = {2018},
  month = {Aug},
  publisher = {American Physical Society},
  doi = {https://link.aps.org/doi/10.1103/PhysRevLett.121.090502},
  url = {https://link.aps.org/doi/10.1103/PhysRevLett.121.090502}
}

@article{qutip5,
  title = {QuTiP 5: The Quantum Toolbox in {Python}},
  author = {
    Lambert, Neill and Gigu{`e}re, Eric and Menczel, Paul and Li, Boxi and
    Hopf, Patrick and Su{'a}rez, Gerardo and Gali, Marc and Lishman, Jake and
    Gadhvi, Rushiraj and Agarwal, Rochisha and Galicia, Asier and Shammah, Nathan and
    Nation, Paul and Johansson, J. R. and Ahmed, Shahnawaz and Cross, Simon and
    Pitchford, Alexander and Nori, Franco},
  journal = {Physics Reports},
  volume = {1153},
  pages = {1-62},
  year = {2026},
  issn = {0370-1573},
  doi = {10.1016/j.physrep.2025.10.001},
  url = {https://www.sciencedirect.com/science/article/pii/S0370157325002704},
}

@article{sarma2025designing,
  title={Designing fast quantum gates using optimal control with a reinforcement-learning ansatz},
  author={Sarma, Bijita and Hartmann, Michael J},
  journal={Physical Review Applied},
  volume={23},
  number={1},
  pages={014015},
  year={2025},
  publisher={APS},
  doi={https://doi.org/10.1103/PhysRevApplied.23.014015},
  url={https://doi.org/10.1103/PhysRevApplied.23.014015} 
}

@article{xiao20222,
	author = {Xiao, Tailong and Fan, Jianping and Zeng, Guihua},
	date = {2022/01/10},
	doi = {https://doi.org/10.1038/s41534-021-00513-z},
	id = {Xiao2022},
	isbn = {2056-6387},
	journal = {npj Quantum Information},
	number = {1},
	pages = {2},
	title = {Parameter estimation in quantum sensing based on deep reinforcement learning},
	url = {https://doi.org/10.1038/s41534-021-00513-z},
	volume = {8},
	year = {2022},
}

@article{PhysRevA110062608,
  title = {Role of bases in quantum optimal control},
  author = {Pagano, Alice and M\"uller, Matthias M. and Calarco, Tommaso and Montangero, Simone and Rembold, Phila},
  journal = {Phys. Rev. A},
  volume = {110},
  issue = {6},
  pages = {062608},
  numpages = {13},
  year = {2024},
  month = {Dec},
  publisher = {American Physical Society},
  doi = {10.1103/PhysRevA.110.062608},
  url = {https://link.aps.org/doi/10.1103/PhysRevA.110.062608}
}

@article{bukov2026reinforcement,
      title={Reinforcement Learning for Quantum Technology}, 
      author={Marin Bukov and Florian Marquardt},
      year={2026},
      journal={arXiv preprint arXiv:2601.18953},
      eprint={2601.18953},
      primaryClass={quant-ph},
      url={https://arxiv.org/abs/2601.18953}, 
}

@article{steveTamingQS,
  title = {Taming Quantum Systems: A Tutorial for Using Shortcuts-To-Adiabaticity, Quantum Optimal Control, and Reinforcement Learning},
  author = {Duncan, Callum W. and Poggi, Pablo M. and Bukov, Marin and Zinner, Nikolaj Thomas and Campbell, Steve},
  journal = {PRX Quantum},
  volume = {6},
  issue = {4},
  pages = {040201},
  numpages = {69},
  year = {2025},
  month = {Oct},
  publisher = {American Physical Society},
  doi = {10.1103/j8c7-v2hd},
  url = {https://link.aps.org/doi/10.1103/j8c7-v2hd}
}

@article{RLPCA2025,
	author = {Fentaw, Haftu W. and Campbell, Steve and Caton, Simon},
	date = {2025/04/26},
	doi = {10.1038/s41598-025-95161-0},
	id = {Fentaw2025},
	isbn = {2045-2322},
	journal = {Scientific Reports},
	number = {1},
	pages = {14605},
	title = {Exploring quantum control landscape and solution space complexity through optimization algorithms and dimensionality reduction},
	url = {https://doi.org/10.1038/s41598-025-95161-0},
	volume = {15},
	year = {2025}
}
%% if required, the content of .bbl file can be included here once bbl is generated
%%\input sn-article.bbl

\end{document}